\documentclass[acmlarge,authorversion,nonacm]{acmart}

\usepackage{booktabs}
\usepackage{multirow}
\usepackage{caption}
\usepackage{siunitx}
\usepackage{graphicx}
\usepackage{booktabs}

\AtBeginDocument{%
  \providecommand\BibTeX{{%
    \normalfont B\kern-0.5em{\scshape i\kern-0.25em b}\kern-0.8em\TeX}}}

\setcopyright{acmcopyright}
\copyrightyear{2023}
\acmYear{2023}
\acmDOI{XXXXXXX.XXXXXXX}

\acmConference[Conference]{2023}
\acmBooktitle{Woodstock '18: ACM Symposium on Neural Gaze Detection,, Woodstock, NY} 
\acmPrice{15.00}
\acmISBN{978-1-4503-XXXX-X/18/06}

\begin{document}

\title{Learning Through AI-Clones: Enhancing Self-Perception and Presentation Performance}



\author{Qingxiao Zheng}
\affiliation{%
  \institution{University at Buffalo}
  \city{NY}
  \country{USA}}
\email{qingxiao@buffalo.edu}

\author{Zhuoer Chen}
\affiliation{%
  \institution{HeyGen}
  \city{IL}
  \country{USA}}
\email{joy@heygen.com}

\author{Yun Huang}
\affiliation{%
  \institution{University of Illinois at Urbana Champaign}
  \city{IL}
  \country{USA}}
\email{yunhuang@illinois.edu}


\begin{abstract}
This study examines the impact of AI-generated digital clones with self-images on enhancing perceptions and skills in online presentations. A mixed-design experiment with 44 international students compared self-recording videos (self-recording group) to AI-clone videos (AI-clone group) for online English presentation practice. 
AI-clone videos were generated using voice cloning, face swapping, lip-syncing, and body-language simulation, refining the repetition, filler words, and pronunciation of participants' original presentations.
Through the lens of social comparison theory, the results showed that AI clones functioned as positive "role models" for facilitating social comparisons. When comparing the effects on self-perceptions, speech qualities, and self-kindness, the self-recording group showed an increase in pronunciation satisfaction. However, the AI-clone group exhibited greater self-kindness, broader observational coverage, and a meaningful transition from a corrective to an enhancive approach in self-critique. 
Moreover, machine-rated scores revealed immediate performance gains only within the AI-clone group. Considering individual differences, aligning interventions with participants' regulatory focus significantly enhanced their learning experience. These findings highlight the theoretical, practical, and ethical implications of AI clones in supporting emotional and cognitive skill development.

\end{abstract}






\maketitle 

\section{Introduction}

A large body of work in the social sciences has investigated the influence of celebrity role models, and in the context of education, several disciplines have a rich research history in this area (e.g. medical education \cite{horsburgh2018skill}). However, in the context of second language learning, research centered on role models has largely remained on the periphery \cite{muir2021role}. Foreign language speakers, especially international students, frequently struggle with unique challenges in online presentation delivery, facing language barriers and difficulties in conveying nuanced messages to global audiences online \cite{moyer2013foreign, szyszka2017pronunciation,  du2015american, reinhardt2019social}. 
Existing technologies such as in situ and delayed feedback systems, although beneficial in providing feedback, can sometimes overwhelm or inadequately serve foreign language speakers, who face unique challenges related to language and cultural differences \cite{wang2020voicecoach, zhou2020effectiveness, moyer2013foreign}. Moreover, these solutions could reinforce "negative performance," posing risks for individuals dealing with public speaking anxiety or self-esteem issues \cite{karl1994will, zhou2020effectiveness}. This situation signifies an urgent need for better {feedback systems}  specifically designed to meet the  challenges of international speakers. In this study, using the lens of \textit{social comparison theory}, we examine how AI can facilitate foreign language speakers in online presentation delivery practice. 

Over 70 years ago,  Festinger \cite{festinger1954theory} introduced the concept of social comparison theory, positing that individuals have an innate drive to evaluate their opinions and abilities for accurate self-assessment. Festinger noted that while people ideally use objective standards for self-evaluation, they often compare themselves to others in the absence of such information. This theory has been extensively applied in education and learning, with numerous studies investigating its effects on learning outcomes \cite{festinger1954theory, dijkstra2008social, micari2011intimidation, hanus2015assessing}. Among the facets explored is the 
\textit{"near-peer role model"}  \cite{murphey1998motivating, michinov2001similarity, kang2019similarity}, which suggests that comparisons are influenced by similarities in gender, race, experience, etc \cite{meisel1990social, wolff2010subjective, luong2020superstars}, indicating a tendency for individuals to compare themselves with those who share similar characteristics. This body of work highlights the complexity of social comparisons and their nuanced impact on personal development and learning.

Many learner tools have been designed and created to support speech and language practice in online environments, offering mechanisms for learners to compare their performance against their peers \cite{guerra2016intelligent, tabuenca2015time}, or against students from previous sessions 
 \cite{davis2016encouraging, perez2017notemyprogress}. However, this method has led to a scarcity of accessible role models, particularly for international or non-native English speakers struggling to find relatable role models because of the lack of representation or connections to others with similar backgrounds \cite{latu2013successful, lockwood2006someone}. Recent advancements in AI technology have made AI-clones increasingly capable, with capabilities ranging from text-based chatbots to immersive virtual reality experiences \cite{lee2023speculating, patel2019ai, freeman2021body}. 
 While these clones present a host of ethical and psychological considerations \cite{stutzman2012boundary, zhao2013many, hogan2010presentation, mori2012uncanny},  their potential benefits, particularly in educational contexts, remain underexplored. One such potential could be the creation of role models \textit{with self-images}, or highly personalized AI-clones, which may support skill enhancement, especially in addressing the distinct challenges faced by international speakers.

In this paper, we propose a novel approach leveraging AI-generated videos to create role models for international speakers practicing online presentations. The proposed intervention is to have users observe their ``improved (potentially ideal) self'' in AI-clones, thereby providing an aspirational target and promoting self-observation from an audience's perspective. The study journey is structured in several key phases: initial self-concept mapping, a first presentation followed by self-assessment, video-stimulated recall for performance review, a second attempt at presentation, and a concluding exit evaluation. Employing a randomized design, we separated participants into a self-recording (control) group and an AI-clone (experiment)  group. We ask,  \textbf{RQ1}: \textit{How do learning outcomes differ between individuals watching self-recorded and AI self-clone videos?}
The AI-clone group was divided into two types of regulatory focus: promotion and prevention. The promotion focus centers on the pursuit of goals and achievements, while the prevention focus emphasizes the avoidance of setbacks and errors~\cite{higgins2001achievement}. We ask, \textbf{RQ2}: \textit{How does regulatory focus influence learning outcomes when using AI self-clone videos?}

To evaluate, we adopt a multi-dimensional lens to examine how self-cloned video feedback can improve presentation efficacy, in the broader context of skill improvements. It involves three stages--\textit{Think}, \textit{Feel}, and \textit{Act}-- to investigate the influence of the AI self-clones on speakers' online presentation practice journeys. This approach is grounded on the concept of self-regulation, which describes the interplay between the environment's impact on an individual and their consequent behavior, mediated through processes of \textit{self-observation}, \textit{self-judgment} or evaluation, and \textit{self-reaction} \cite{teng2022can, bandura1986social}.

This study contributes to the existing body of research on AI-supported online presentation delivery in three ways. 
First,  we demonstrate that AI-generated self-clone videos can also function as ``role models,'' in line with the social comparison and regulatory focus theory, without the need to explicitly prime a role model like previous studies \cite{latu2013successful, lockwood2006someone, zhang2016can, lockwood2002motivation}. 
Second, our research delves into the effects of generative AI self-clones on learners' self-observation, perceptions of their speech and learning experiences, and their performance. Participants using the AI exhibited enhancements particularly in the emotional aspects (e.g., showing kindness to self) and self-perception (e.g., perceived higher improvements in smiles and communication) in a presentation practice learning context \cite{orii2022designing, pataranutaporn2022ai}.
Our findings indicate that AI's influence varies with learners' regulatory focus. Promotion-focused individuals experienced increased pronunciation confidence, reduced speech anxiety, and found AI more enjoyable and useful, unlike their prevention-focused counterparts. While both groups improved in reducing filler words, only those with a prevention focus also saw enhancements in managing repetition and weak words. These insights open new avenues for applying AI in online speech and language learning.

\section{Related Work}

\subsection{Urgent Needs and Approaches for Practicing Online Self-Presentation}
The digital age, marked by the ubiquity of social media platforms like Facebook, TikTok, Instagram, Youtube, and LinkedIn, has underscored the importance of effective online self-presentation, or in the perpetual act of curating their online persona \cite{jacobson2020you, schlosser2020self}. This has become particularly crucial for enhancing social and professional opportunities \cite{chan2019becoming, li2021tell}. The widespread adoption of remote work practices and platforms such as Zoom, Teams, and Webex has made online self-presentation a necessary skill for social interaction and work proficiency \cite{gunawan2021application}. However, international students face unique challenges in online self-presentation for non-native speech \cite{moyer2013foreign}, including but not limited to, language issues such as accents and pronunciation \cite{szpyra2014pronunciation, szyszka2017pronunciation}, vocabulary \cite{hazenberg1996defining, roberts2005misunderstandings}, grammar \cite{du2015american}, and the capability to convey nuanced messages that resonate with a global audiences \cite{du2015american, reinhardt2019social}.

\subsubsection{\textbf{Limitations of Current Feedback Systems}}
Existing feedback systems significantly advance the domain of public speaking and online presentation training, they mainly fall into three categories: feedback systems—comprising in-situ, delayed, and self-corrective feedback. 

In-situ feedback systems provide real-time cues that improve speakers' awareness of nonverbal cues and pacing. {For instance, these systems can deliver live feedback on body openness, energy, and speech rate to help users improve their nonverbal behaviors \cite{damian2015augmenting}. They can also assist speakers by providing immediate prompts, such as showing texts of \textit{“louder”} or \textit{“faster”} in AR environments, enabling adjustments to volume and speaking rate \cite{tanveer2015rhema, barmaki2018embodiment}. Some systems display script coverage to aid in oral presentations \cite{asadi2017intelliprompter}.} However, real-time feedback can overwhelm international speakers who may already face challenges related to language and cultural nuances. Moreover, these systems often lack tools to facilitate deliberate practice or performance comparisons \cite{wang2020voicecoach}.

Delayed feedback systems usually offer quantitative metrics that guide future improvements \cite{wang2020voicecoach}.  These include assessment, analytical, or tagged “videotape playback” tools such as dashboards visualizing performance metrics or physiological responses like electrodermal activity (e.g., heart rate) linked to stress and anxiety \cite{croft2004differential, giraud2013multimodal}. Automatic performance scoring systems assess speech quality using vocal characteristics like pitch, speech rate, and loudness \cite{wang2020voicecoach}. While such systems support novice users in reflecting on their performance \cite{zhou2020effectiveness}, they often rely on pre-defined, context-insensitive metrics. This approach can be particularly limiting for international speakers unfamiliar with the rhetorical or cultural norms of the presentation language \cite{wang2020voicecoach, zhou2020effectiveness}.

Self-corrective feedback systems, such as auditory feedback mechanisms, provide immediate remedies for presentation issues, and also widely use playback as a mechanism. For example, computer-aided personalized pronunciation training with exaggerated corrective feedback improves speakers’ awareness \cite{bu2021pteacher}. Altered auditory feedback systems resynthesize voice inputs with modifications to create confident-sounding voices, helping speakers release tension and manage emotions \cite{naruse2019estimating, orii2022designing}. Despite their effectiveness for short-term error correction, intermittent feedback has been shown to have limited impact on long-term retention \cite{schmidt1989summary, wang2020voicecoach}.

A major drawback of the above feedback systems is their tendency to highlight negative performance, also referred to as “negative performance” \cite{karl1994will}. For example, In-situ video feedback systems have been reported to distract speakers by focusing attention on their imperfections  \cite{dermody2019recommendations}.
Similarly, video or audio playback in delayed feedback systems and self-corrective feedback systems frequently exacerbates public speaking anxiety \cite{breen1970effect}. Individuals with public speaking anxiety, who could benefit from playback feedback, often avoid it due to self-esteem and self-efficacy issues \cite{karl1994will, zhou2020effectiveness}.  These issues can be particularly pronounced for international speakers, who may interpret negative feedback as criticism of their linguistic or cultural proficiency. To overcome this limitation, it is crucial to explore feedback mechanisms that focus on positive reinforcement for international foreign language speakers. 

\subsubsection{\textbf{Challenges of Learning from Role Models}}

Role modeling is widely proven as an effective mechanism for improving self-presentation skills, allowing learners to observe and internalize strategies for success \cite{orii2022designing}.  Observing peers, lecturers, or exemplary speech performances has been shown to enhance self-presentation efficacy and encourage informative communication \cite{schunk1987peer, sonnenschein1980development}. To leverage the role modeling mechanism, prior work mostly used "\textit{priming}" to set role models, by priming (picturing) a role model who is distinct from oneself, yet shares similarities, to achieve their impact in many of the learning contexts \cite{zhang2016can,lockwood2002motivation, lockwood2006someone}. However, there is a lack of availability of "real human role models" who can be accessed on-call. 

Research studies also indicate that role models who share attributes with learners are more effective in enhancing learning outcomes \cite{latu2013successful, lockwood2006someone}. These near-peer role models are defined as “people who might be “near” to oneself in several ways, such as age, ethnicity, sex, interests, past or present experiences \cite{murphey1998motivating}, which is often referred to in second language learning context. For example, prior research highlights the impact of near-peer role models on learners' perceptions of speech performance \cite{latu2013successful,  lockwood2006someone}. Similarly,  gender plays a significant role in the effectiveness of role models, with a study reporting women achieving better self-evaluations and presentation outcomes when observing female role models with similar attributes\cite{latu2013successful}. 

For international speakers or non-native English speakers, however, finding role models who share similar attributes can be challenging, particularly when situated in a foreign country where the majority of their connections and interactions are with native speakers, making it difficult to identify relatable role models.  This disconnect can hinder their ability to fully engage with or imitate the role models, reducing the effectiveness of the priming approach \cite{orii2022designing}. Despite its promise, this approach has not yet been widely integrated into the design of feedback systems. 

The limitations of existing feedback systems, particularly their tendency to reinforce negative performance, together with the lack of resources to identify relatable role models, motivate us to think: If and how can the role-model approach be adapted to feedback systems to provide presentation feedback that avoids negative reinforcement while enhancing reliability?

\subsection{The Potential of Using AI-clones as  Role Models}
An AI-clone refers to the digital replication of an individual's appearance and behavior, made increasingly feasible through recent advancements in deep-learning technologies such as DeepFake, voice conversion, and virtual avatars \cite{lee2023speculating, westerlund2019emergence}. These AI-clones manifest in various formats, ranging from straightforward text-based chatbots \cite{patel2019ai} to immersive virtual reality experiences that include multi-modal feedback mechanisms \cite{freeman2021body}. AI-clones are created by training machine learning algorithms with data specific to a person, including their vocal characteristics and visual appearances, to emulate their speech and behavior \cite{maras2019determining}. As a result, AI-clones can convincingly replicate or even refine the unique traits of real people \cite{fletcher2018deepfakes}. This makes it feasible to create a role model using an AI-clone that mirrors speaker-related characteristics while enhancing specific aspects of their performance, such as pronunciation or voice.

Existing studies on AI-clones have primarily focused on the negative implications of deepfake technologies. However, recent research has begun exploring their potential role in learning contexts, particularly in simulating peers or instructors. For instance, studies on AI-generated virtual instructors based on admired individuals have demonstrated their ability to boost student motivation and foster positive emotions, even though they had no significant impact on test scores \cite{pataranutaporn2022ai}. These findings suggest that AI-clones, particularly those tailored to individual needs, could provide a unique and \textit{personalized} educational experience. Additionally, AI co-presenters in virtual presentations have been found to reduce anxiety and improve presentation quality, especially among participants who perceive the agents favorably \cite{kimani2021sharing, kimani2019you}. These findings highlight the potential for AI-clones to serve as role models or co-presenters, offering benefits to learners facing public speaking anxiety or challenges like linguistic and cultural barriers.

Despite these advances, these studies have focused on using AI-clones to simulate others rather than employing them as self-representations or self-images for skill development. Given the growing accessibility and popularity of this technology, there is an urgent need to understand how personalized self-images created by AI-clones can benefit learning. In addition, 
while AI-clones offer promising opportunities, they also present challenges, such as issues of impression management \cite{stutzman2012boundary, zhao2013many, hogan2010presentation}, identity and authenticity concerns \cite{michikyan2015can, pringle2015conjuring, devito2017platforms}, and the uncanny valley effect \cite{mori2012uncanny}, which can provoke unease in users encountering digital replicas of themselves\cite{weisman2021face, mori2012uncanny}. These complexities raise important questions about the ethical application of AI-clones and their potential pitfalls in learning contexts.

By addressing these challenges and leveraging the unique capabilities of AI-clones, this research aims to explore the potential and pitfalls of using AI-clones as a personalized feedback system that supports learning while fostering positive reinforcement.

\subsection{Theoretical Lens: Social Comparison Theory and Individual Regulatory Focus}

Role models serve as key reference points for social comparison, a psychological process described by Social Comparison Theory, which posits that people have an innate drive to evaluate themselves by comparing their performance to others \cite{festinger1954theory}. In learning contexts, role models significantly influence perceptions, recognition, and productivity, shaping the efficiency of learning experiences and outcomes \cite{sjerps2010bounds, oppenheim2010dark, sheldon2008priming}. Social comparison can take two forms: upward comparison, where individuals compare themselves to others who perform better, and downward comparison, where they compare themselves to those performing worse \cite{fleur2023social, dijkstra2008social, micari2011intimidation, hanus2015assessing}. 

Within this framework, the concept of \textit{Regulatory Focus} provides a deeper understanding of how individuals respond to social comparison information. Regulatory Focus identifies two motivational orientations: \textit{“promotion focus”}, which emphasizes the pursuit of aspirations and accomplishments, and \textit{“prevention focus”}, which centers on avoiding failures and mistakes \cite{higgins2001achievement}. These orientations determine whether upward or downward social comparisons are more inspiring and motivating to an individual \cite{zhang2016can, watling2012understanding, rosenzweig2016you}

Studies have shown that promotion-focused individuals are motivated by \textit{“upward comparisons”}, drawing inspiration from role models who exemplify success and highlight pathways to achievement \cite{pasko2021roles, lockwood2002motivation, lockwood2006someone}. Conversely, prevention-focused individuals are more inspired by \textit{“downward comparisons”}, as they emphasize avoiding mistakes and minimizing risks \cite{lockwood2008impact, higgins2001achievement}. 
This alignment between an individual’s regulatory focus and the type of social comparison or role model is referred to as congruence, while a mismatch is termed incongruence \cite{higgins2001achievement}. 
For example, Fleur et al. demonstrated that a learning analytics tool displaying students' performance alongside peers with similar goal grades fostered extrinsic motivation and improved academic achievement, aligning comparisons with students’ regulatory focus \cite{fleur2023social}. In the context of learning, promotion-focused individuals are most motivated by positive role models who showcase strategies for success. Meanwhile, prevention-focused individuals benefit from negative role models who emphasize avoiding mistakes. 

To evaluate the effectiveness of AI-clones as role models for international speakers in online presentation practice, we adopt the concept of {\textit{“upward comparison”}} by presenting AI-clones as positive role models. 
Upward comparison is relevant here as the task focuses on inspiring learners to improve, aligning with goals of skill development and aspirational growth, unlike downward comparison, which emphasizes avoiding negative outcomes  (e.g., staying above a minimum threshold or avoiding below-average scores). Moreover, this focus addresses the “negative reinforcement” prevalent in current feedback systems \cite{karl1994will, zhou2020effectiveness},  ensuring the emphasis remains on learners' potential rather than their shortcomings. Thus, \textbf{H1} examines the comparative effect of AI-clone and the traditional playback approach used in mainstream feedback systems. Recognizing that the impact of social comparison varies based on an individual’s regulatory focus, we also investigate how promotion-focused and prevention-focused learners may respond differently to AI-clone. This leads us to \textbf{H2}, which explores individual differences in using AI-clone for presentation practice.

\subsubsection{{\textbf{H1 (self-recording group v.s. AI-clone group)}}}
To examine the comparative effect of AI clones and the traditional playback approach used in mainstream feedback systems, we developed several hypotheses based on previous studies on online speech and language training.

\paragraph{Self-perception}
Supporting speakers in improving their self-perception of speech is crucial, especially since enhanced self-perceptions have a positive effect on social well-being \cite{pettersson2018psychological}. 
Specifically, recent research in speech and language training has investigated how listening to recordings of role-model speakers affects self-perception of speech, and researchers found that exposure to model speakers significantly improved confidence in tone \cite{orii2022designing}, when measuring four aspects of self-perceptions on \textit{confidence, satisfaction, communication, and self-expression}. Building on these findings, we aim to compare the effectiveness of using different role models in enhancing speech perception metrics. We hypothesize, {\textbf{H1 (a)}: \textit{Using AI-clone videos as a role model for speech practice will be more effective than using self-recording videos in improving self-perception of speech performance.}}

\paragraph{{Speech Quality}}
Moreover, while the effects of role models in fostering self-perceptual changes are studied  \cite{orii2022designing, nash2016if, scherer1973voice}, the direct impact of role models on immediate behavioral changes in fluency, pronunciation, or delivery was also explored in some studies \cite{orii2022designing, pataranutaporn2022ai}. Thus, we hypothesize, {\textbf{H1 (b)}:  \textit{Using AI-clone videos as a role model for speech practice will be more effective than using self-recording videos in improving speech quality.}}

\paragraph{Self-kindness}
Additionally, addressing the issue of negative reinforcement in current feedback systems is crucial. Previous studies evaluated the self-kindness \cite{shepherd2009negative}  which inquired about their self-response to observed mistakes or imperfections. We want to understand if the AI-clone can alleviate the "negative reinforcement" issue previously discussed \cite{karl1994will, zhou2020effectiveness}. By presenting an aspirational version of the self, AI-clones may foster self-kindness, encouraging learners to reflect positively on mistakes and imperfections, Thus, we hypothesize, {\textbf{H1 (c)}: \textit{Using AI-clone videos as a role model for speech practice will be more effective than using self-recording videos in improving self-kindness.}}

\subsubsection{\textbf{H2 (AI-clone group: Promotion focus v.s. Prevention focus)}}
{Second, to examine the influence of individual differences on the effectiveness of AI-clone in speech practice, we employed constructs related to regulatory focus within educational contexts.}

\paragraph{Learning Experience}
Previous research has demonstrated that aligning an individual's regulatory focus with the traits of their role models positively impacts their assessment of the \textit{learning experience} \cite{jarvela2023predicting,zhao2024exploring,reinhold2021students}. 
Previous research has thoroughly investigated how the alignment between an individual's regulatory focus and the traits of their role models can positively influence their assessment of the learning experience \cite{jarvela2023predicting,zhao2024exploring,reinhold2021students}. It has been found that learners' experience is considered often boosted when their role models have similar goals and reduced when the goals differ \cite{lockwood2002motivation, zhang2016can, jiang2022motivation}. 
A prior study found that when a role model is aligned with learners' regulatory focus, it increases learners' \textit{motivation} to learn and their evaluation of the learning tool as helpful, but they typically do not yield positive evaluations regarding the enjoyment \cite{zhang2016can}. Another study also reported that adopting a promotion-focused approach has been found to reduce \textit{anxiety} in foreign language practice contexts \cite{jiang2022motivation}, an aspect which is critical in presentation \cite{nash2016if}. Building on this foundation, we hypothesize, {\textbf{H2 (a)}: \textit{Using AI-cloned videos as a role model for speech practice, individuals with a promotion focus will demonstrate higher evaluation on the learning experience. }}

\paragraph{Self-perception} Building upon the broader rationale established in H1, where role models are expected to enhance self-perception, we extend our investigation into how regulatory focus moderates these effects, we further hypothesize, {\textbf{H2 (b)}: \textit{Using AI-cloned videos as a role model for speech practice, individuals with a promotion focus will demonstrate higher improvements in self-perception of speech performance.}}

\paragraph{Speech Quality}
Building upon the broader rationale established in H1, where role models are expected to speech performance, we extend our investigation into how regulatory focus moderates these effects, we further hypothesize, \textbf{H2c}: \textit{Using AI-cloned videos as a role model for speech practice, individuals with a promotion focus will demonstrate higher improvements in speech quality, compared to those with a prevention focus. }

\subsubsection{\textbf{{Measurements}}}
{We present details on measurement including the pre-test of regulatory focus in Table \ref{measurements}. We calculated the mean scores of the items for the multi-dimensional variables and reported the reliability.}

\begin{table}[H]
\small
\renewcommand{\arraystretch}{1.5} 
\resizebox{\columnwidth}{!}{%
\begin{tabular}{p{2cm}p{2cm}p{6cm}p{3cm}}
\hline
\textbf{Areas} &
  \textbf{Variables} &
  \textbf{Sample Items} &
  \textbf{Notes} \\
\hline

\multirow{4}{=}{\textbf{Self-perception \cite{orii2022designing}}} &
  Confidence &
  I am satisfied and fulfilled with my voice and the way I speak. Confidence is the ability to effectively perform a task. &
  \multirow{4}{=}{7-point Likert scale: 1 = strongly disagree to 7 = strongly agree; same statement of each item used for rating for all 9 aspects of speech qualities; collected at 2 time-points} \\
 &
  Satisfaction &
  I am satisfied with my voice and the way I speak. Satisfaction refers to the fulfillment of happiness, i.e., “Are you happy with what you are evaluating?” &
   \\
 &
  Communication &
  I can communicate what I want to say or convey to the listener, through my voice and the way I speak. &
   \\
 &
  Self-Expression &
  I can express myself through my voice and the way I speak. &
   \\
\hline
\multirow{2}{=}{\textbf{Speech qualities \cite{scherer1973voice}}} &
  Word choice &
  Repetition, filler words, and weak words, along with their respective percentages. &
  \multirow{2}{=}{Presented as statistics; collected at 2 time-points; analyzed by Yoodli} \\
 &
  Delivery &
  Eye contact percentage, pacing in words per minute, pacing variation, minimum/maximum pause duration, mean pause duration. &
   \\
\hline
\textbf{Self-kindness \cite{shepherd2009negative}} &
  ~ &
  Was your response to yourself regarding the mistakes or imperfections observed in your video positive, negative, or neutral? &
  Single-choice \\
\hline
\multirow{4}{=}{\textbf{Learning experience}} &
  Motivation \cite{lockwood2002motivation} &
  I want to integrate AI videos’ knowledge with my own knowledge for online speech practice. &
  \multirow{4}{=}{7-point Likert scale: 1 = strongly disagree to 7 = strongly agree; 4-item for each variable; Cronbach's alpha ranges from 0.84 to 0.94. 
  } \\
 &
  Helpfulness \cite{zhang2016can} &
  I think learning from AI videos to improve my online speech was helpful. &
   \\
 &
  Enjoyment \cite{zhang2016can} &
  I think learning from my AI video was a happy experience. &
   \\
 &
  Anxiety \cite{bartholomay2016public} &
  AI videos helped me manage concerns like, "My speech won't impress the audience." &
   \\
\hline
\textbf{Regulatory Focus Questionnaire (RFQ)} \cite{higgins2001achievement} &
  ~ &
  I feel like I have made progress toward being successful in my life. (Promotion) How often did you obey rules and regulations that were established by your parents? (Prevention) &
  11 items using 5-point scales with scoring keys \\
\hline
\textbf{Manipulation Check} &
  ~ &
  How would you rate the performance of the AI-clone for your speech practice, compared to your performance? &
 7-point scale ranging from 1 (AI's performance is much worse) to 7 (AI's performance is much better) \\
\hline
\end{tabular}%
}
\caption{{Measurement areas, variables, sample items, and scales used. Note: Yoodli is a third-party, time-stamped, automated delayed speech analysis feedback generating tool on word selection and delivery nuances for presentation assessment used in prior research studies \cite{whalen2023my, panke2023workshop}.}}
\label{measurements}
\end{table}

\section{Method}

\subsection{Multi-Modal AI-clone Manipulation } 
\begin{figure*}[ht!]
    \centering
    \includegraphics[width=0.9\textwidth]{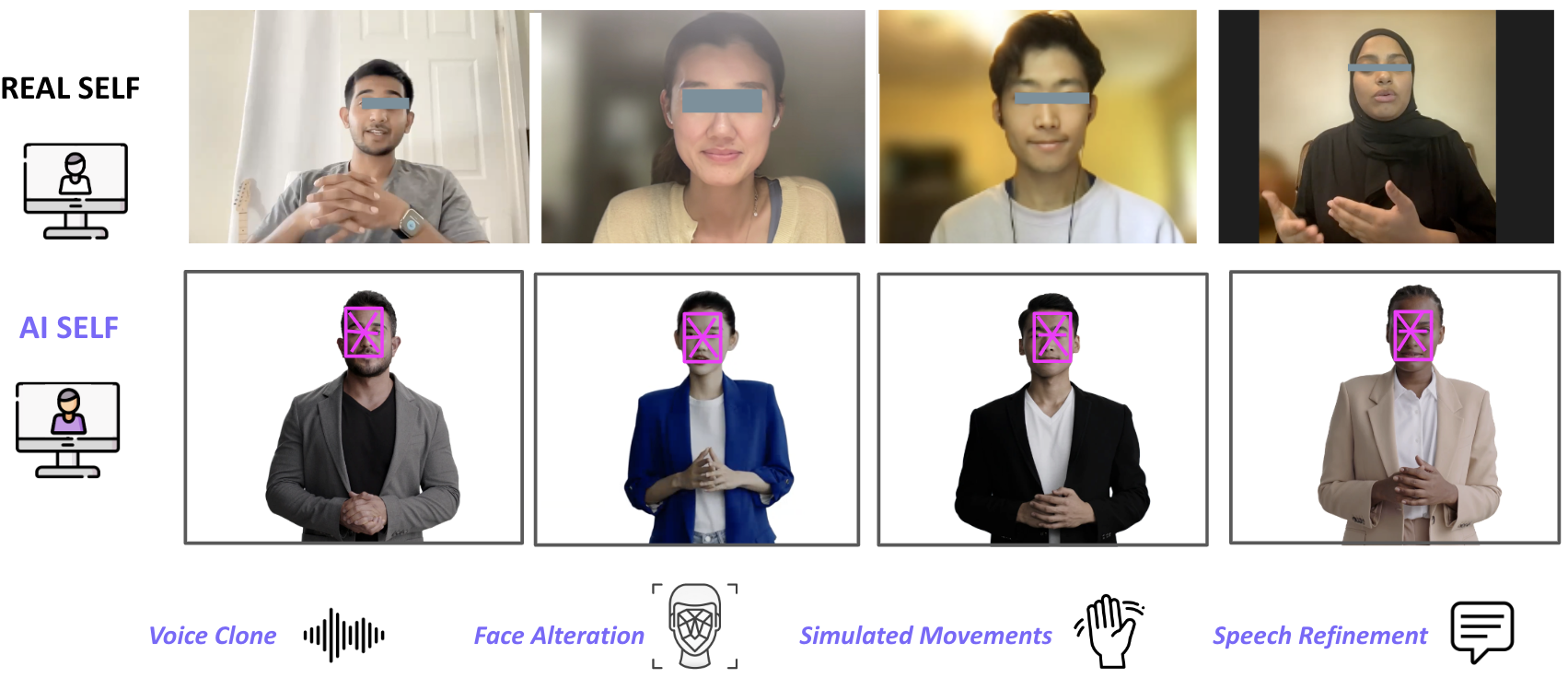} 
    \caption{AI-clones with participants' chosen Avatar template (Images are shared with participants' permission)}
    \Description{Figure 1 shows samples of screenshots of participants' real-self and AI-self, as well as the technologies used to generate AI-self. The first row of images are sample screenshots of participants' real-self, and the second row of images are sample screenshots of participants' AI-self.}
    \label{AIvideos}
\end{figure*}

To create an AI-clone or digital replication that is programmed to exhibit the behaviors, language skills, and presentation styles that the learner aims to achieve, providing a positive stimulus, we gathered a two-minute audio sample and multiple snapshots from each participant to serve as the input data for training our models. The goal of this training was to accurately replicate both the vocal and facial characteristics of the participants in the generated videos, as illustrated in Figure \ref{AIvideos}, a comparison of \textit{self-recording} and \textit{AI-clones} videos.

The AI-clone is generated through a three-step process. First, we leverage the text-to-speech eleven-labs toolkit\footnote{https://elevenlabs.io} to capture the unique vocal traits of participants and create AI-generated audios \cite{11labs}. Second, we refine the transcripts from participants' initial presentations by removing repetitive and filler words. These optimized transcripts are then used to produce synthetic audio that features a Standard English Accent but still utilizes the participants' original vocal characteristics. {Lastly, we use Heygen\footnote{https://www.heygen.com}, a commercial platform that generates customizable AI avatars to replicate participants' facial talking and gestures \cite{heygen}, which are captured through snapshots. This synthesized audio is then paired with a participant-chosen preset avatar template, and face-swapping techniques are employed to generate AI-clones of the participants.}

The manipulation check result (M= 6.69, SD= 0.79) indicates the AI-clones can be a positive role model in general, see measurement in Table \ref{measurements}.

\subsection{Study Procedure}

\begin{figure*}
    \centering
    \includegraphics[width=0.9\textwidth]{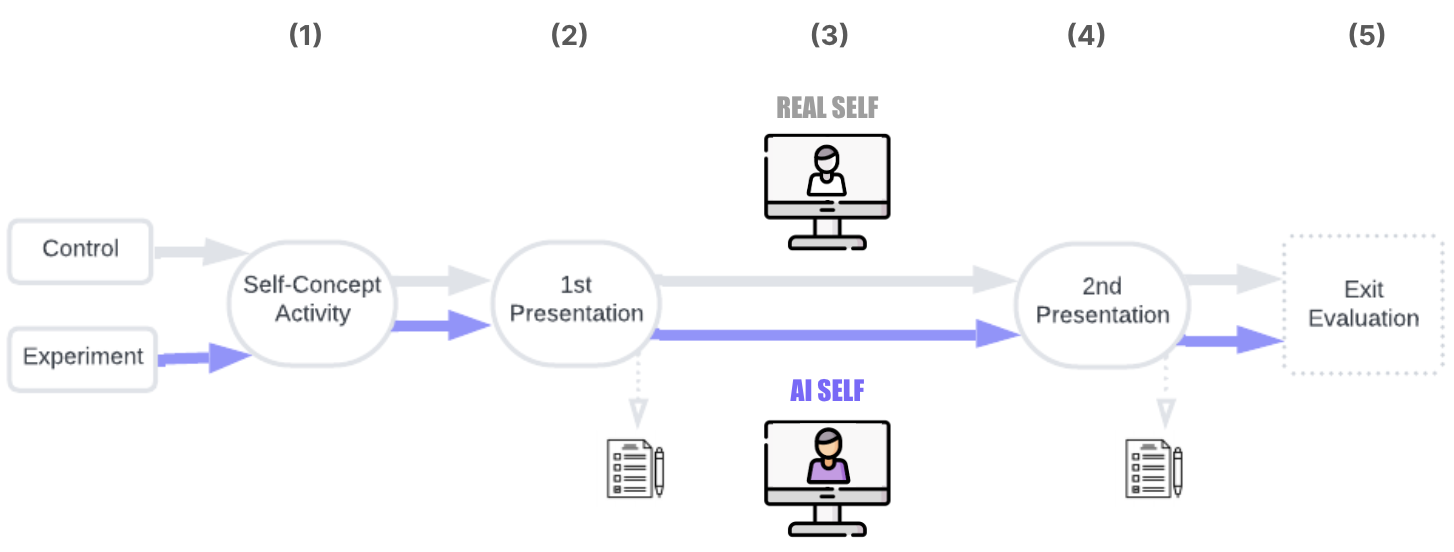} 
    \caption{Study procedure: all participants engaged in the five steps (detailed in Section 3.2).}
    \Description{Figure 2 shows the procedure of this study, which is as follows:
    1. Participants are recruited and assigned to the experiment group and control group randomly.
    2. Participants listed 5-10 attributes that represented their "Actual Self".
    3. Participants perform the first presentation and then finish a self-assessment.
    4. Participants watch a sample video as needed. 
    5. Participants perform the second presentation and then finish a self-assessment. 
    6. Participants finish an exit evaluation.}
    \label{fig: procedure}
\end{figure*}

The experiment is approximately 100 minutes long and follows the study procedure in figure \ref{fig: procedure}. Before the formal experiment, they read a sample script and recorded a 2-minute video for AI-clone generation. Then, all participants engaged in the following sequential activities.

\begin{enumerate}
    
    \item \textit{\textbf{Self-Concept:}} To warm up, participants listed 5-10 attributes that represented their \textit{"Actual Self"} (e.g., "I tend to repeat myself in online presentations"). They also described their envisioned self -- an \textit{"Ideal Self"} for promotion-focused individuals (e.g., "I want to be persuasive") or an \textit{"Ought-to-Self"} for prevention-focused individuals (e.g., "I must maintain consistent eye contact"). This information was documented in a shared online file. Then, participants selected an outfit from a template gallery that aligned with their online speech.

    \item \textbf{\textit{The 1st Presentation: }}Participants had 10 minutes to prepare for a one-minute short speech in English, to concisely summarize an idea, project, product, or oneself for online business, networking, interviews, or quick essence conveyance -- without relying on a script. The speech was recorded for later analysis of presentation quality. Afterward, they finished a questionnaire asking about self-perception of speech performance. 
    
    \item \textbf{\textit{Video Replay:}} A stimulated recall \cite{lyle2003stimulated} is a form that allows participants to think-aloud their thoughts while reviewing a video interpretation. Participants first watched a sample video showing the process and were then asked to share their screen as they played back their video, with the option to replay as needed. The AI-clone group viewed AI-clone video based on their initial speeches, while the self-recording group watched replays of their presentations. After this, participants assessed whether the video is a positive role model for manipulation check.

    \item \textbf{\textit{The 2nd Presentation:}}  Reflecting on their self-concept, participants \textit{wrote their goals} for their second presentation and were allotted 10 minutes for preparation. Upon completion of their second one-minute speech, they finished a questionnaire asking about self-perception of speech performance, and evaluation on their learning experience.

     \item \textbf{\textit{Exit Evaluation}}: For the exit evaluation, participants discussed their experience watching their video playbacks (either self-recording or AI-clone). Then, they watched a counterpart video— either AI-clone or self-recording, depending on their group—and repeated the step (3). Finally, they were interviewed about using AI videos for learning, its benefits and drawbacks. 
 
\end{enumerate}

\subsection{Data Analysis}
The socio-cognitive view of human behavior describes self-practicing as the reciprocal determinism of the environment on the person mediated through behavior (self-observation, self-evaluation, and self-reaction) \cite{teng2022can, bandura1986social}. Based on this view, we adopt a holistic evaluation involving {{Think}}, {{Feel}}, and {{Act }} to explore how AI-clones influence speakers' presentation practice experience.

\subsubsection{\textbf{Quantitative Data}}
We employed statistical approaches to evaluate the result for between-group analyses. To ensure a robust evaluation of training outcomes, we selected statistical methods based on the timing and nature of the collected data.

\paragraph{Two Time-Point Measures (Pre-Post).} When data were collected both before and after the practice, we used paired sample t-tests to examine within-group changes across various dependent variables. This method allows us to analyze how each group's performance changed due to the practice. Additionally, to control for initial performance levels and isolate the impact of starting points, we employed the Analysis of Covariance (ANCOVA). This "post-adjust between" approach is particularly effective for making comparative assessments between the two groups in randomized pre-post designs where outcomes are measured both at baseline and after intervention \cite{wan2021statistical}. For ANCOVA, we conducted post-diagnostic plots to confirm the linearity and homogeneity of regression slopes, and residual analysis indicated normal distribution and equal variances in these variables, thereby meeting the data assumptions.

\paragraph{Single Time-Point Measures.} When data were collected or coded only at a one-time point, independent t-tests were applied to assess differences in continuous outcomes between the groups. This "between" approach allows us to compare group means for each dependent variable at that specific time point. If the outcomes are categorical with small sample sizes, Fisher's exact tests were utilized to evaluate differences between groups \cite{fisher1970statistical, teachey2022children}. This method provides a more accurate assessment of group differences for non-continuous variables. 

\subsubsection{\textbf{Qualitative Data}}
Transcripts provided insights into participants' cognitive processes as they interacted with the AI system. Through iterative coding and constant comparison, we captured the nuances of participants' opinions, challenges, and suggestions.

\paragraph{Think-Aloud Transcripts}
Transcripts have been gathered for the think-aloud (stimulated recall) sessions of each participant when replaying videos. All transcriptions were then carefully reviewed. An experienced researcher, along with two undergraduate research assistants, independently applied and tallied codes to the segmented transcripts based on the predefined coding scheme, which covered various dimensions participants remarked on, such as eye contact, pace, and pronunciation. Subsequent analysis focused on the frequency of these specific codes, referred to as "number of reflections," and the diversity of dimensions discussed, termed "dimension counts." This process enabled the identification of patterns and trends among the groups.

\paragraph{Written Goals}
Participants wrote goals for the second presentation.  First, by reading through their goals, we identified that all participants exhibited some level of attention to the following topics: {\textit{technical}} proficiency (e.g., pause, word choice, pace, grammar, time, content, pronunciation), and {\textit{emotional}} resonance in delivery (e.g., eye contact, tone, gesture, emotion, facial expression), echoed with the previous work on two dimensions of evaluating presentation \cite{de2012effective}. We also found participants, used different verb choices: the {\textit{correction words}} that show errors, e.g., \textit{"avoid," "don't," "fix," "improve"}, and the {\textit{enhancement words}} that show efforts, e.g.,  \textit{"add," "more," "should," "try"}. Then, we quantified the frequency of participants' disclosures tied to these four categories. We compared the differences between technical proficiency and emotional resonance, and correction approach and enhancive approach, and used a binary coding (Y/N) to determine if participants demonstrated a stronger focus for either aspect. These results were then organized into a 2x2 contingency table and proportions were analyzed.

\paragraph{Semi-Structured Interviews}
In the semi-structured interviews, 
participants were guided to express their experience, presentation preparation challenges, and insights on integrating AI videos for presentation skill improvements, identifying AI benefits, proposing integration strategies, and addressing concerns. Interviews were audio-recorded and transcribed through Zoom.

\subsection{Participants}
\subsubsection{Recruitment} We recruited participants through university-hosted events organized by the International Student and Scholar Services (ISSS) and connections established through various social media platforms. Participants were randomly selected based on their responses to a screening survey. Booking invitations were extended to individuals who met the following criteria: 1) Non-native English speakers, 2) Individuals hoping to improve online presentations in English, 3) Age exceeding 18 years. Participants received a copy of consent before the study and our research was approved by the IRB. Participants received \$15 of compensation upon the completion of the study.

\subsubsection{Randomization} 
At the start of the study, participants were asked to complete an initial registration survey that collected information on their age range and native language. In addition, they were required to fill out the \textit{"Regulatory Focus Questionnaire"} \cite{higgins2001achievement} to assess their regulatory focuses. Using this information, the study implemented a randomized design for dividing participants into two primary groups, each ideally comprising 15 participants. However, due to the absence of one participant from the self-recording group, the allocation was slightly altered, resulting in the self-recording group (control, \(n=14\) and the AI-clone group (experiment, \(n=15\)), bringing the total count in these initial groups to 29 participants.

To explore how individual regulatory focus might influence the impact of AI-clone videos, the study further divided the AI-clone group based on participants' regulatory focuses. This division aimed to form two additional groups, each reflecting a balanced representation of the regulatory focuses to ensure statistical integrity. Stratified random sampling was utilized for this purpose, drawing from the remaining pool of registrants. This approach facilitated the creation of two groups: the promotion-focused group (\(n=15\)) and the prevention-focused group (\(n=15\)). The final participant count in the two new groups was 30 participants. Consequently, this segmentation increased the total participant count for this study to 44 participants.

\subsubsection{Demographics}
Of our participants, 45.24\% are female and 54.76\% are male. Participants' age ranges from 18-24 (44.2\%) to 25-34 (55.8\%). For their native languages, the majority are Mandarin, making up approximately 52.38\% of the total. Indians are the second most common, making up approximately 9.52\%, followed by Koreans at 11.36\%.  All other native languages make up 29.55\% of the sample, including Bangladeshi, Cantonese, Ecuadorian, French, Hispanic/Latino, Indonesian, Japanese, Jordanian, Kyrgyz, Nepali, Nigerian, Peruvian, and Vietnamese.

\section{Findings}
\subsection{\textbf{Differences Between Self-Recording and AI-Clone Groups (RQ1)}}

\subsubsection{{\textbf{H1 (a), rejected.}}}
The self-recording group demonstrated significantly higher self-perception of satisfaction in pronunciation than the AI-clone group. 

By comparing self-perceptual measures (i.e., confidence, satisfaction, expressiveness, and communication) at two-time points, the ANCOVA showed that when the baseline is considered, significant differences \textit{between-group} were observed for the satisfaction in pronunciation: \( F(1, 26) = 10.50, p = 0.003 \), effect size: \( \eta^2 \)  = 0.257, large effect. 
The self-recording group showed a higher adjusted mean (\(\textit{adjusted mean} = 5.70\)) than the AI-clone group (\(\textit{adjusted mean} = 4.53\)) in satisfaction with pronunciation, thus H1a was rejected.  See Figure {\ref{fig:satisfaction}}, and the complete result in Appendix Table.  {\ref{Ancova-CE}}. 

A follow-up analysis of \textit{within-group} changes was also conducted, see details of the paired-sample t-tests in Appendix Table. {\ref{Paired-t-test-CE}}. 
Results showed that: 1) both groups exhibited statistically significant improvements in overall confidence,  confidence in pace and eye contact, overall satisfaction, satisfaction in fluency, eye contact, pace, and pauses, and expressiveness; 2) The self-recording group showed unique improvements in both confidence and satisfaction of pronunciation (\(t = -4.32\)), and confidence of Loudness (\(t = -2.88\)); and 3) the AI-clone group showed unique improvements in {communication} (\(t = -2.48\)), satisfaction of {repetition} (\(t = -2.75\)), confidence of managing {pauses} (\(t = -2.96\)), and confidence of {smiles} (\(t = -2.44\)).

\begin{figure*}
    \centering
    \includegraphics[width=1\linewidth]{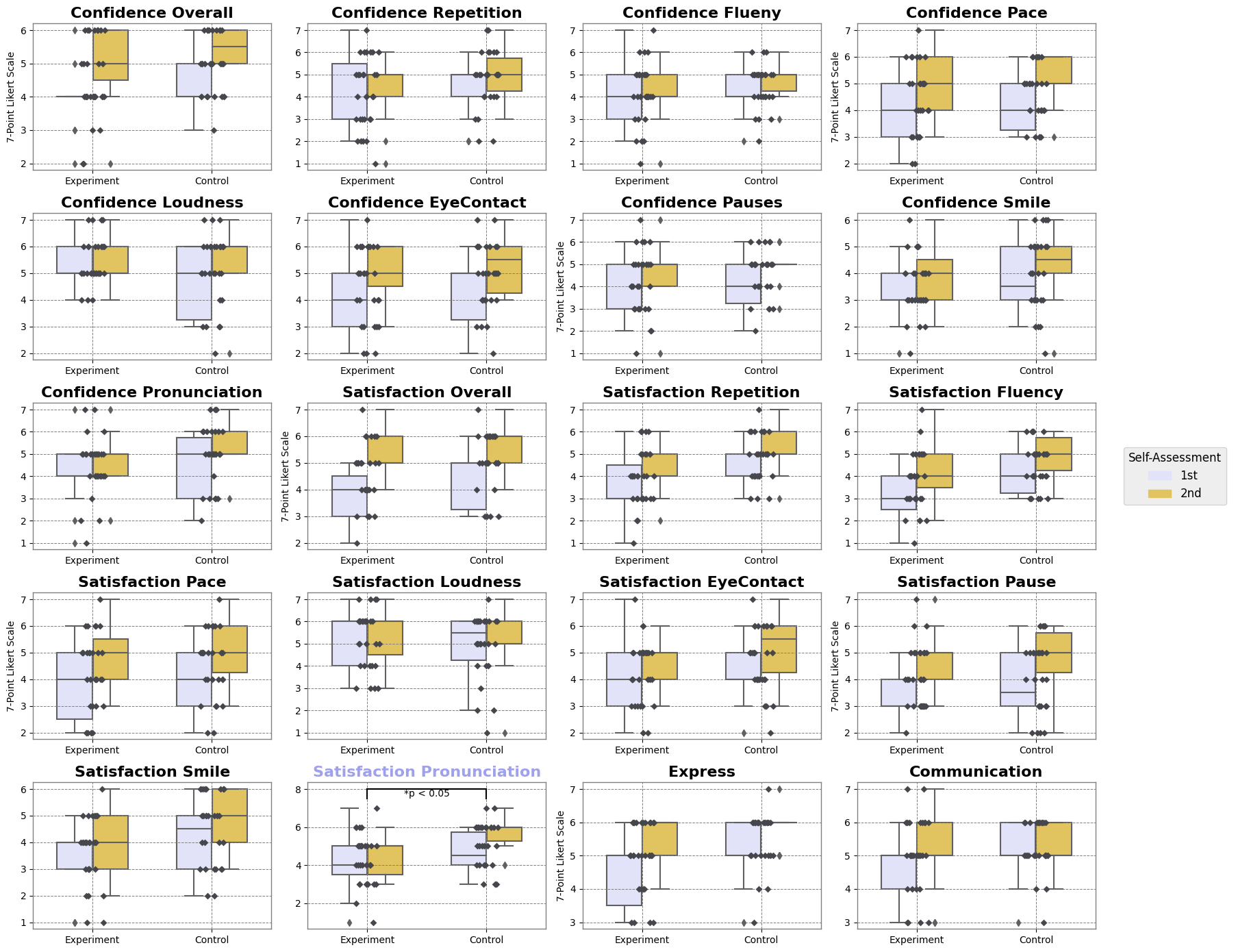}
    \caption{{Pre-test and Post-test \textit{self-perceptions} towards the rated 9 items of speech performance. While many improvements were found within both groups, only \textit{Satisfaction in Pronunciation} yielded a significant difference between groups. }}
    \Description{Figure 3 shows the boxplots of confidence and satisfaction metrics differences between the experiment group and the self-recording group based on ANCOVA results. After adjusting for the participants' 1st self-assessment, the self-recording group showed a significantly higher satisfaction level in "pronunciation" in the 2nd self-assessment compared to the AI-clone group.}
    \label{fig:satisfaction}
\end{figure*}

\subsubsection{{\textbf{H1 (b), partially supported. }}} The AI-clone group demonstrated a significant immediate improvement in speech quality of pace control than the self-recording group.

\begin{figure*}
    \centering
    \includegraphics[width=1\linewidth]{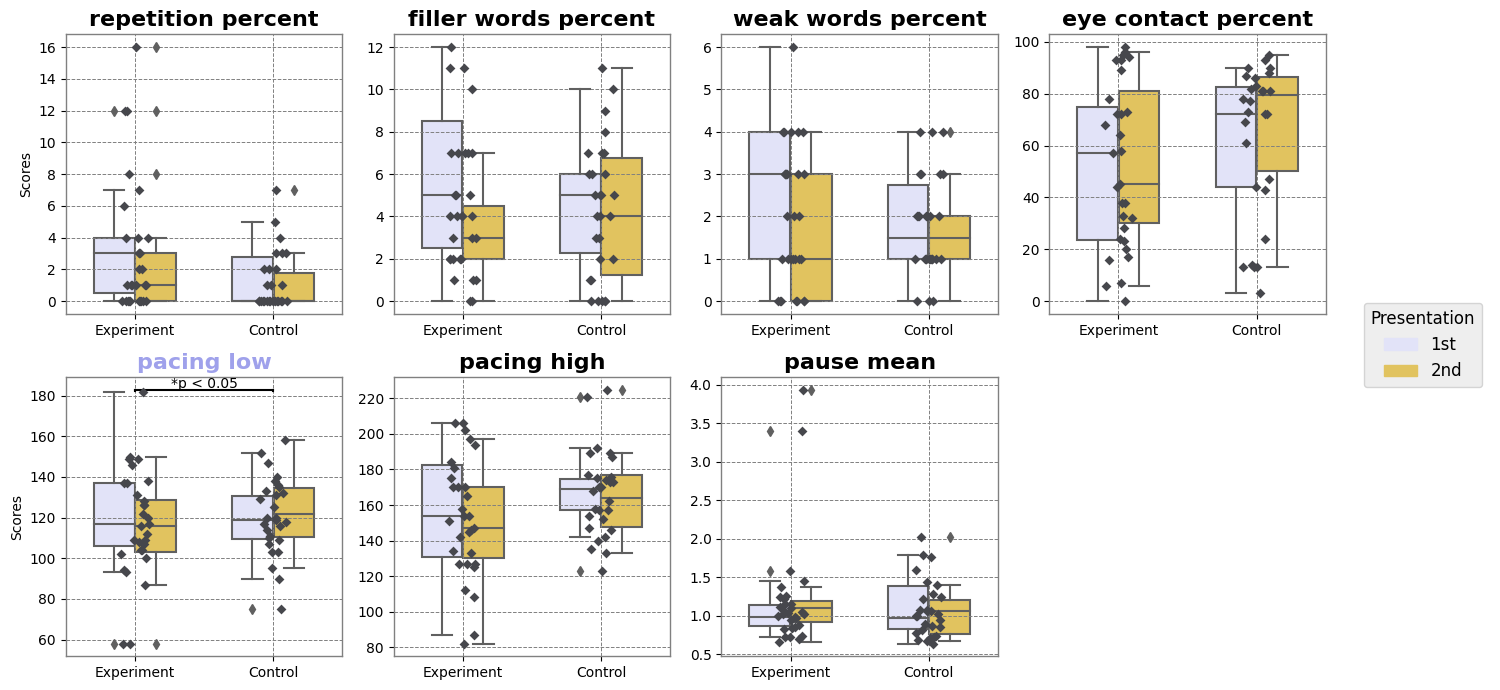}
    \caption{{Pre-test and Post-test Scores of \textit{Speech Quality} for the Self-Recording and AI-Clone Group: Only \textit{pacing control} yielded a significant difference between groups. 1st is original, 2nd is after trained. }}
    \Description{Figure 4 shows the boxplots of performance metrics differences between the AI video group and self-recording group based on ANCOVA results. After adjusting for participants’ 1st presentation scores, the self-recording group showed a significantly higher satisfaction level in "pacing (low)" in the second presentation compared to the AI video group.}
    \label{fig:yodii}
\end{figure*}

ANCOVA showed that significant effects between the self-recording and the AI-clone group were observed for controlling pace: \( F(1, 26) = 4.50, p = 0.044 \), effect size: \( \eta^2 \)  = 0.04, small to medium effect. The covariate-adjusted means revealed that the AI-clone group (\textit{adjusted mean} = 114.93 w/m) reported a lower speech speed compared to the self-recording group (\textit{adjusted mean} = 122.71 w/m) during their second performance after adjustments to their baseline performance, thus {H1 (b)} was partially supported. Post-diagnostic plots and residual analysis indicated the data met the assumptions. See details in Figure \ref{fig:yodii} and Appendix Table. {\ref{Ancova-CE}}. 

A follow-up analysis of \textit{within-group} changes was also conducted, see details of the paired-sample t-tests in Appendix Table. {\ref{Paired-t-test-CE}}. Results showed that: Only immediate speech performance improvements were found in the AI-clone group, whereas the self-recording group found no actual immediate speech improvement. Specifically, the AI-clone group exhibited significant reductions in filler words (\(t = 3.10\)), weak words (\(t = 3.07\)), and their respective percentages (\(t = 3.03, t = 2.78\)), all with \(p\)-values below 0.05. In contrast, within the self-recording group, results showed no significant differences in these performance metrics.

\subsubsection{{\textbf{H1 (c), supported.}}}
The AI-clone group showed higher self-kindness than the self-recording group.
Fisher's exact analysis revealed a statistically significant difference in the likelihood of negative self-ratings (p < 0.05). Individuals in the self-recordings group were significantly more likely to rate themselves negatively (Odds Ratio: 26.0, 95\% CI: 2.61 to 259.29) compared to those in the AI-clone group. This statistical significance (p-value: 0.0017) suggested a strong association between self-recording feedback and an increased likelihood of negative self-assessment, compared to AI-generated feedback. Thus, H1 (c) was supported. 

\textbf{Summary:} 
No significant differences were observed between groups in self-perception measures, except for the self-recording group exhibited significantly greater improvements in pronunciation satisfaction compared to the AI-clone group. Follow-up within-group analysis reinforced this finding, as participants in the self-recording group showed unique enhancements in both confidence and satisfaction with pronunciation, a trend not seen in the AI-clone group. 

Meanwhile, the AI-clone group uniquely improved in self-perception in communication, repetition, pauses, and smiling, which were not observed in the self-recording group. Also, the AI-clone group was significantly more inclined to extend self-kindness. Moreover, participants' actual speech quality only revealed enhancements in the AI-clone group, especially in word choice and pacing control. 

\subsection{\textbf{Differences Within the AI-clone Group (RQ2)}}

We further explored the influence of AI-clone videos to assess potential individual differences based on regulatory focus, involving participants from both promotion-focused and prevention-focused groups.

\subsubsection{{\textbf{H2 (a), partially supported. }}}
Promotion-focused individuals reported higher enjoyment, helpfulness, and anxiety reduction than prevention-focused individuals. No significant differences were found for motivation, thus, {H2 (a)} was partially supported. These differences are detailed in Table  \ref{experience}, with statistical evidence provided by the independent t-test results with medium to large effects.

\begin{table}[ht]
\centering
\label{table:t-test-results}
\begin{tabular}{lcccccccc}
\toprule
& Prevention (n=15) & Promotion (n=15)& t & p \\
\midrule
Enjoyment    & 5.43±1.23 & 6.33±0.67 & -2.486 & 0.019*        \\
Helpfulness  &  5.48±1.19 & 6.30±0.52 & -2.431 & 0.022*    \\
Less Anxiety  &  4.73±0.96 & 5.73±0.89 & -2.966 & 0.006**  \\
Motivation    & 5.68±1.36 & 6.20±0.54 & -1.368 & 0.182        \\
\bottomrule
\end{tabular}
\caption{Independent T-test Evaluation of Learning Experience}
\label{experience}
\end{table}

\subsubsection{{\textbf{H2 (b), partially supported.}}}
Promotion-focused individuals exhibited significantly higher confidence in pronunciation than prevention-focused individuals.

ANCOVA analysis indicated significant differences \textit{between-group}  in perceived confidence of  {pronunciation} (\(F(1, 27) = 5.434, p = 0.027\)), with a large effect size (\(\eta^2 = 0.168\)). Promotion-focused participants reported higher confidence levels (\(\textit{adjusted mean} = 5.40\)) compared to their prevention-focused counterparts (\(\textit{adjusted mean} = 4.53\)) in their second speech, highlighting a notable difference between the groups. Thus, {H2 (b)} is partially supported. 
See Appendix Table. {\ref{Ancova-PP}} for details.  

A follow-up analysis of \textit{within-group} changes was also conducted, see details of the paired-sample t-tests in Appendix Table.{\ref{Paired-t-test-PP}}. 
We found: 1) both groups showed improvements in overall satisfaction, satisfaction in repetition, satisfaction in pace control, satisfaction in eye contact, satisfaction in pauses, and expressiveness; 2) The promotion-focused participants showed unique improvements in 7 aspects: overall confidence, confidence in repetition, confidence in eye contact, confidence in pauses, confidence in pronunciation, satisfaction in fluency, satisfaction in pronunciation; and 3) the prevention-focused participants demonstrated unique improvements in 3 aspects: communication, confidence in fluency, confidence in pace.

\subsubsection{{\textbf{H2 (c), rejected. }}}
Prevention-focused individuals had more immediate improvement dimensions than promotion-focused individuals. 

ANCOVA found no significant \textit{between-group} differences in immediate speech performance, thus, {H2 (c)} was not supported.  See details in Appendix Table. {\ref{Ancova-PP}}. 

A follow-up analysis of \textit{within-group} changes was also conducted, see details of the paired-sample t-tests in Appendix Table. {\ref{Paired-t-test-PP}}. 
We found:  1) both groups had a decrease in filler words and filler words percentage; and 2) unique to the prevention group, there was a significant decrease in repetition and weak words, both in terms of count and percentage.

\textbf{Summary:} Promotion-focused individuals exhibited significantly more reductions in speech anxiety, along with increased enjoyment and perceived usefulness when using AI-clone video for presentation practice, compared to prevention-focused individuals. Also, they reported significantly higher perceived confidence in pronunciation. However, no significant difference in immediate speech performance was found. Both promotion-focused and prevention-focused individuals improved in reducing filler words, but only prevention-focused individuals improved in controlling repetition and weak words. 

\subsection{More Insights from Qualitative Analysis}
Using the data collected from participants' \textit{Think-Aloud Transcripts}, and \textit{Written Goals}, we found that participants in the self-recording and AI-clone groups diverged in their observations, priorities, and approaches.

\subsubsection{\textbf{AI-Clone Videos Encouraged More Coverage of Self-Observations Than Self-Recordings}}
Using the transcripts from the video think-aloud, we coded the participants' mentioned dimensions. The independent t-tests were conducted to compare the differences between groups watching self-recording and the AI-clone videos. We found that the dimensions observed by AI-clone group (M = 1.33 SD = 1.53) were statistically higher than the self-recording group  (M = -0.32, SD = 2.55), t = -2.133, p < 0.05, Cohen's d = 0.78, medium effect. The results suggested the AI-clone group was more adept at identifying and reflecting upon various nuanced dimensions of their speech performances (e.g., repetition, filler words, and eye contact) than their counterparts in the self-recording. 

Participants also expressed their desire to emulate specific styles or qualities that they admire or find effective in the AI-clones. We found explicit mention of AI, e.g., \textit{"I want to pace myself like the AI (excluding the first introduction). I would like to do gestures {similar to the AI} and also pronounce some words like it. "} (P16). This statement not only recognizes a certain style as effective but actively seeks to incorporate elements of that style into the presenter's style. Speakers also extracted valuable lessons from AI. For example, while some participants recognized the AI's limitation in terms of energy, they were inspired to combine the AI's other strengths with their own desired improvements. For example, \textit{"I think I will try to use a more energetic voice than AI and make my speech more engaging and fun" }(P32). This reflects a nuanced approach to self-improvement.

\subsubsection{\textbf{Goal Setting Indicates Different Priorities for Emotional Resonance}}

For the written goals for the second delivery, both the self-recording and the AI-clone groups emphasized the importance of technical proficiency and emotional resonance in presentation skills.
Some participants primarily targeted the  "nuts and bolts", or foundational elements of a presentation—focusing on fluency (P21, P23, P25, P26, P28, P29, P30, P42), pacing (P24, P37, P42, P43), vocabulary (P25, P28, P37, P42), grammar (P30), less repetition (P21), pronunciation (P24), time control (P24), content (P22, P26, P27, P44) and minimizing filler words(p26). Their goals were inherently technical, striving for a well-tuned presentation that was polished in its clarity, form, and structure. For example, one participant said, \textit{"Control my speaking pace. Don't speak too fast and give an unclear speech"} (P24), while another aimed to \textit{"Speak fluently, use fewer filler words, think and speak simultaneously, and diversify my vocabulary"} (P25).

Some participants aimed for a well-rounded presentation experience that extended beyond technical prowess to include emotional resonance and audience engagement. They set goals for technical aspects such as fluency (P14, P16, P18, P32, P34, P36), word choice (P34), pacing (P16, P31, P38), content (P14, P33, P35), pronunciation (P16), and minimizing filler words (P16, P35, P36, P38). However, some of them also mentioned utilizing {logical pauses for emphasis} (P16), and more also prioritized body language (P14, P15, P16, P31, P33), {facial expression} (P18, P20, P31, P39, P41),  eye contact (P17, P18, P32, P34, P35, P39, P41),  tone (P32), and {emotional expressiveness} (P15, P32, P41) to not just to present information but to make it resonate on a more emotional to forge a deeper connection with the audience. For example, participants prioritize emotive expression to connect with the audience, such as \textit{"I want to {appear} to be more emotional by adding some gestures during speaking"} (P15); \textit{"be more confident, making eye contact {with audiences} better, smiling, energetic, and emotional"} (P41).

Further fisher's exact test comparing the goal-setting tendency within each of the self-recording and AI-clone indicated a significant statistical difference in delivery-related tendencies (emotional resonance v.s. technical proficiency). The odds of reporting a emotional outcome were significantly lower in the self-recording group compared to the AI-clone group (odds ratio = 0.08, 95\% CI: 0.008 to 0.85, p< 0.05). Notably, even though the AI-clone group observed more observations during their think-aloud sessions than the self-recording group—as the finding in the think-aloud observation—they still laid significantly more emphasis on emotional resonance aspects for their presentations goal. The findings suggest that participants watching the AI-clone recording are significantly more likely to aim goals in emotional aspects compared to those in the self-recording group. This could be attributed to the perceived feedback provided by AI, which may provide them an alternative perspective. However, given the small sample, future research should explore this association in larger, more diverse populations to evaluate the strength of this association.

\subsubsection{\textbf{Linguistic Traits Indicating Different Approaches to Self-Critique}}

We also coded the two types of linguistic traits (corrective and enhancive) participants used. Some participants took a corrective stance, aiming to rectify perceived weaknesses or limitations in their existing skill set. The aim is to bring the presentation up to a certain standard by eliminating errors or gaps in proficiency and adhering to an outline, among other things. For example, participants were concerned with minimizing mistakes to improve the overall quality of the presentation: \textit{"I should stick to my outline of what to say in paper so I don't take too many pauses and don't look like I am unprepared or don't know what to say"} (P29),  \textit{"avoid the repeated words and pause if possible" }(P21), and \textit{"Use some complex vocabs if possible"} (P28). The underlying assumption here is that repeating words, taking unnecessary pauses, or using not complex vocabulary is {flaws} that need to be corrected for a more polished presentation to avoid {weaknesses}. Some participants, on the other hand, demonstrated a seemingly more enhancive approach, where the focus seems not merely on fixing shortcomings but on elevating already-existing competencies --  focusing on amplifying and expanding existing skills. These goals of adding emotional layers, utilizing gestures, and regulating the pace more effectively amplify the strengths of a presentation by adding new layers of complexity and engagement. For example, \textit{"Use {one more} example or illustrations, and maybe add a bit of hand movement  {when appropriate}"} (P14), and \textit{"Deliver more useful information" }(P20). These are not about correcting flaws but about enriching the content to provide added value to the audience. The implicit understanding is that the basic presentation is already solid; what's needed is the “extra mile” to make it more engaging or more informative. 

Fisher's exact test was conducted between the self-recording and the AI-clone groups on the average number of the two approaches (enhancive v.s. corrective) each group took. The result indicated a significant statistical difference in goal-setting approaches between the self-recording and the AI-clone (odds ratio = 0.11, 95\% CI: 0.017 to 0.72, p< 0.05), indicating the self-recording group is more oriented toward corrective measures, aiming to fix existing weaknesses, while the AI-clone group leans more toward enhancive strategies, seeking to build on existing strengths, although both groups engage in a mix of goal-setting approaches. However, given the small sample, future research should explore this association in larger, more diverse populations to evaluate the strength of this association. 

\textbf{Summary:} The qualitative data reveal variations between the groups that observed self-recordings and those that viewed AI-clone videos. These differences are evident in the coverage of observations, goal-setting priorities, and self-critique approaches. The findings illuminate how using AI-clones as underlying mechanisms (e.g., observation attention, linguistic traits) helps identify and discuss these "unconscious" nuances and concerns associated with their use.

\section{Discussion}
Our study offers a multi-dimensional exploration of the impact of using AI-clone for online presentation practice and reveals several important {theoretical, practical, and ethical} implications of leveraging AI clones in self-presentation contexts.

\subsection{Advancing Social Comparison Theory: New Insights on AI Clones as Role Models}
{Our findings contributed to the empirical understanding on social comparison theory in the context of using AI clones of oneself as social comparison entities, a field which has traditionally focused mainly on human role models or using others as role models  \cite{festinger1954theory, dijkstra2008social, micari2011intimidation, hanus2015assessing}.} In today's digital landscape, where online presentations are prevalent across various platforms—from workplace virtual conferences to self-promotion on social media \cite{pasquini2021being}, for foreign language speakers engaged in self-directed online presentation practice, the quest for relatable role models poses a significant challenge. We found that using AI-clone could potentially serve as an accessible positive role model when practicing online self-presentation.

On one hand, in our investigation, AI-clones were utilized as a means for upward comparison in presentation practice learning. To the best of our knowledge, this research is the first to explore applying \textit{social comparison theory} \cite{festinger1954theory} to the use of one's self-image as a primed role model, marking a departure from previous approaches that used “others” as role models \cite{latu2013successful, lockwood2006someone, zhang2016can, lockwood2002motivation}. This method of self-comparison does not rely on external cues or suggestions to identify comparison targets, which could potentially introduce bias or affect the authenticity of the comparison  \cite{ratcliff1995bias}. Our manipulation check confirmed that the AI clone was indeed perceived as positive, and this is not a self-evident result because individuals tend to perceive positive role models as similar to themselves, rather than superior \cite{suls2013handbook, schokker2010impact}. This was evidenced by manipulation check ratings deviating from the scale’s neutral midpoint, indicating participants viewed their AI-clones' performance as surpassing their original performance. Findings extend existing literature that emphasizes the effectiveness of relatable role models in learning \cite{latu2013successful, lockwood2006someone}. 

On the other hand, we investigated the hypothesis that an individual's regulatory focus moderates the influence of role models in contexts featuring positive role models. Our findings validate the theory that alignment between an individual's regulatory focus and their goals leads to increased learning engagement in the new AI-clone context, evidenced by significant upticks in \textit{enjoyment} and perceived \textit{helpfulness} of the learning process. This observation is consistent with previous research examining regulatory focus theory within online learning environments \cite{zhang2016can}. Additionally, we noted a significant reduction in \textit{anxiety}, a vital factor for effective presentation practice, echoing findings from studies but applied different approaches to role modeling in learning and training contexts \cite{jiang2022motivation, leitzelar2020regulatory}. These insights advance our understanding of how individual differences mediate the effectiveness of AI-based feedback.

\subsection{{Identifying the Practical Potentials of AI Clones for  Self-Presentation Training}}
We discussed the practical implications of using AI clones for self-presentation training as follows. 

\subsubsection{{Four Benefits of Adopting AI Clones Identified}}

\paragraph{{A. Immediate Improvements in Speech Quality (H1-b).}}
H1-b was partially supported, as the AI-clone group demonstrated measurable enhancements in speech pace and word choice, particularly through the reduced use of filler and weak words. These results highlight the potential of AI clones to deliver immediate, actionable feedback that refines technical aspects of speech. This finding complements prior research focused on perceptual changes, offering a more granular understanding of AI’s impact on immediate speech performance improvements \cite{orii2022designing, pataranutaporn2022ai}.

\paragraph{{B. Promoting Self-Kindness and Counteracting Negative Reinforcement (H1-c).}}
Significant findings showed that the AI-clone group exhibited higher levels of {\textit{self-kindness}} compared to the self-recording group. This suggests that AI-clone videos may mitigate the "negative reinforcement" often associated with traditional playback systems \cite{karl1994will, zhou2020effectiveness, dermody2019recommendations}. Previous studies reported that self-compassion techniques, including self-kindness, help athletes view mistakes as learning opportunities, reducing the negative impact of self-criticism and enhancing well-being\cite{miyuki_omorimiyake_2021}. This transformative approach positions AI clones as valuable tools for enhancing both performance and emotional well-being in learning contexts. Future studies could explore the mechanisms underlying the observed increase in self-kindness, such as the role of reliability, or deeper knowledge about how it may foster acknowledging errors without self-condemnation, allowing for constructive reflection and growth as discussed by previous work \cite{kristin_d__neff__2014, kristin_d__neff_2022}.

\paragraph{{C. Fostering Engagement Skills (H1-a, Follow-Up Analysis).}}
The AI-clone group demonstrated within-group improvements in engagement skills, such as communication, pauses, and smiling, areas that did not improve in the self-recording group. These results suggest that structured feedback from AI clones provides specific insights and cues that are difficult to identify through self-recording. This aligns with AI’s role in enhancing self-reflection and engagement, helping learners refine their naturalness and expressiveness during presentations.

\paragraph{{{D. Enhancing Self-Reflection and Emotional Resonance (Qualitative Insights).}}}
Qualitative analysis revealed that the AI-clone group engaged in deeper \textit{self-reflection}, emphasizing not only technical skills but also \textit{emotional resonance} and \textit{enhancive strategies}. AI clones enabled learners to explore subtle or "unconscious" elements of their performance that traditional playback methods often overlook. By providing objective feedback, AI clones helped learners recognize both their unique \textit{strengths} and areas for improvement, potentially fostering autonomy and competence. These findings align with Self-Determination Theory (SDT), which highlights autonomy, competence, and relatedness as key drivers of motivated behavior \cite{deci2012self}. Additionally, comparing oneself to an AI-enhanced "ideal self" supports autonomy by empowering learners to control their learning trajectory and connect their current abilities to future potential.

\paragraph{Advantages of Self-Recording on Pronunciation (H1-a) and Discrepancies Between Perception and Performance (H1-b).} H1-a was rejected, as the self-recording group demonstrated significantly higher satisfaction with pronunciation compared to the AI-clone group. This advantage likely stems from the direct feedback mechanism inherent to self-recording, where participants closely listen to and critique their own pronunciation and volume. This reflective process facilitates awareness of these two aspects of speech, highlighting the unique benefits of self-recording for enhancing awareness of these aspects. While previous studies focused on perceived confidence in tone using recordings \cite{orii2022designing}, our findings reveal that self-recording offers new advantages for pronunciation satisfaction and technical skill refinement. Interestingly, a notable discrepancy emerged between self-perception and objective performance metrics: While self-recording participants reported higher satisfaction with pronunciation, pace, and pauses, their performance metrics revealed faster speech and shorter pauses, potentially signaling anxiety rather than genuine improvement. This underscores the importance of integrating objective feedback mechanisms, such as those offered by AI assessment tools, to complement self-assessment and provide a more balanced understanding of performance.

\subsubsection{Impacts of Individual Regulatory Focus When Using AI Clones for Training}  

\paragraph{Regulatory Fit Benefits More in Learning Experience (H2-a).}
Promotion-focused individuals reported a significantly more positive learning experience when using AI-cloned videos, demonstrating higher levels of enjoyment and perceived helpfulness of the AI tool, along with a notable reduction in anxiety related to speech performance. These findings align with prior research on regulatory fit \cite{higgins2001achievement}, which suggests that a promotion-focused orientation—centered on seeking gains—paired with a positive role model enhances the subjective value and emotional benefits of learning systems. This regulatory alignment makes the experience more engaging and emotionally supportive for promotion-focused learners.

\paragraph{Regulatory Fit Demonstrated Higher Self-perceptions (H2-b).} Promotion-focused individuals also exhibited significantly greater perceived confidence in their pronunciation compared to their prevention-focused counterparts.  This aligns with the tendency of promotion-focused individuals to respond positively to aspirational and success-oriented feedback, which AI clones are designed to provide.

\paragraph{Enhanced Speech Quality for Prevention-Focused Individuals (H2-c).}
The result suggested that while the promotion and prevention had no differences in performance improvements when comparing their first and second presentations, the prevention-focused participants had immediate improvements in some speech metrics that the promotion-focused group did not show. Specifically, prevention-focused participants showed immediate improvements in specific speech quality metrics, such as reducing filler words, controlling repetition, and minimizing weak words. These improvements were not observed in the promotion-focused group, reflecting the prevention focus's orientation toward avoiding losses and errors \cite{higgins2001achievement, zhang2016can}. This immediate enhancement highlights how AI-based learning systems can cater effectively to prevention-focused individuals by emphasizing precision and error reduction. Given that H2 (c) was not supported, this finding, unreported in prior work, prompted us to hypothesize that AI's most powerful impact might be in ‘‘shaping perception" rather than in delivering ‘‘immediate performance improvements". Nevertheless, perceptual gains should not be undervalued, as they may be the precursors to long-term performance improvements \cite{williams2005perceptual,schouten2007precursors}.

\subsubsection{{Bringing inspirations for future research on self-representation}}
{Previous studies have demonstrated the effectiveness of role models in various training contexts, including but not limited to social skills \cite{ladd1983cognitive}, career development \cite{lamb2022impact}, therapy \cite{rochlin1982sexual}, leadership, and medical training \cite{spaans2023role}.
For example, they can be integrated to support individuals in addressing challenges like social anxiety issues with role-playing exercises, such as practicing conversations or presentations with different versions of their self-images, without the pressure of real-world interactions. This aligns with findings that highlight the effectiveness of virtual reality and simulated agents in reducing anxiety and improving therapeutic outcomes \cite{horigome2020virtual, wong2023systematic}. 
By providing contextually relevant feedback and simulating realistic reactions based on a self-image, AI clones have the potential to support these contexts while fostering self-reflection. }

{Building on our findings of four key benefits of AI clones—enhancing speech quality, promoting self-kindness in the face of errors, improving engagement skills in interpersonal communication, and fostering emotional resonance—we envision future research further exploring their role in developing personal self-representation skills. Researchers could also examine additional variables relevant to specific contexts or explore potential negative effects on self-image, such as those discussed by Farrar et al. \cite{farrar2015self}. These effects might include an overemphasis on transforming the self to align with the perceived ideal characteristics of the AI clone, potentially leading to unrealistic self-perceptions or reliance on external validation \cite{kang2009exploring}.
These areas of inquiry are essential for future studies to deepen our understanding of the broader implications and applications of AI clones.}

\subsection{Ethical Considerations}

{We reported various concerns based on interview studies, addressing issues related to personal identity and selfhood, concerns about excessive complexity, expectations for alignment in human-AI interactions, and anxieties about inadvertently presenting inconsistent self-images. For detailed qualitative insights, see Appendix \ref{ethics}.}

Besides the widely discussed ethical considerations surrounding privacy, data security, informed consent, and the potential misuse of AI-generated clones, e.g., the abusive potential of AI, the displacement of self-identity \cite{haimson2015online}, and negative emotions \cite{lee2023speculating}, we posit that any AI technology should respect individual differences: Ignoring individual differences in AI design can inadvertently violate ethical obligations. A single approach may not address the unique learning needs, cultural sensitivities, and emotional responses of different users (e.g., individual regulatory focus). This is especially true for international speakers who already face linguistic and cultural barriers (e.g., preferences for maintaining unique accents even though they are not \textit{"standard"}). Tailoring AI-clone to individualized needs is not just a matter of educational efficacy; it becomes an ethical obligation to ensure equitable access to educational opportunities.

Our findings also point out that people have varying emotional and cognitive responses to AI technologies. Designers must take these psychological factors into account to avoid unintended negative consequences. For instance, while AI-clone could alleviate public speaking anxiety for some, they might exacerbate feelings of inadequacy or self-consciousness for others, e.g., \textit{Feels too good to be true"} (P41). The potential for negative psychological impact becomes even more acute in educational settings (e.g., while most participants react positively to AI, we noted that two participants mentioned being \textit{"Uncomfortable to face the AI because it copies me "(P37)} and "\textit{I don't like hearing my voice} (P14)". Previous literature \cite{weisman2021face} has shown users to be uncomfortable with seeing their personality and likeness being embodied in interactive applications. With the advent of AI-clones of even-increasing fidelity, issues of impression management \cite{hogan2010presentation, zhang2021social} and uncanny valley perceptions \cite{draude2011intermediaries, weisman2021face, welker2020trading} are problems that will have to be resolved to successfully develop applications that use AI-clones.

The ethical deployment of AI-clone in educational settings should also consider accessibility. AI-driven solutions often require advanced software, potentially excluding those who cannot afford or access these resources, e.g., based on our case, training AI requires high-quality audio, image, or sometimes video data, however, this becomes a barrier for those who cannot afford advanced devices. Additionally, the focus on high-quality audio-visual data for model training may marginalize specific groups, such as those who are blind or have speech impairments. Inclusivity means designing for the widest range of users, and this principle should be adhered to when deploying AI-clone in educational contexts. A more extensive focus on the ethical issues that might arise is necessary as a future direction.

\section{Conclusion}
This study leverages advanced AI technology to create AI-clones, offering a novel approach to self-training for online presentations. These AI-clones address limitations in current feedback systems, such as negative reinforcement, while serving as self-improvement tools and culturally sensitive role models. Emphasizing the importance of personalization based on individual regulatory focus, the findings offer significant theoretical, practical, and ethical insights for future research and applications.

\section{Limitation and Future Research} The study also comes with some limitations. We primarily recruited an international student sample, representing the younger generation, who are generally more receptive and trusting towards technology, as suggested by prior work \cite{wu2022exploring}. While the external validity of our findings might be examined in future studies, it is important to note that we do not claim that our findings are applicable to a wider population throughout this paper. Secondly, due to the cost of creating AI-clones with emerging technology, the sample size of this study is limited. However, we have carefully addressed and reported the effect size of our findings, allowing future studies to build on our current findings and associations to further validate the results. Thirdly, the focus of our research was on the immediate effects of interacting with AI-clones, particularly in the context of presentation delivery training. The study, as indicated by its title and content, does not explore the long-term sustainability of the observed behavioral changes. Given that mastering presentation skills requires time, subsequent longitudinal research could delve into whether initial perceptual improvements lead to enduring performance enhancements and whether improvements are maintained over time.




\bibliographystyle{ACM-Reference-Format}
\bibliography{2-REFERENCES.bib}

\appendix

\section{Statistic Analysis Results}

\begin{table}[H]
\centering
\begin{tabular}{lrlrl}
\toprule
                                          Variable &  T (AI-Clone) & Significance &  T (Self-Recording) & Significance\\
\midrule
                    word\_count\_p1 vs word\_count\_p2 &            1.69 &                        No &         1.10 &                     No \\
                    repetition\_p1 vs repetition\_p2 &            1.42 &                        No &        -0.17 &                     No \\
    repetition\_percent\_p1 vs repetition\_percent\_p2 &            0.00 &                        No &         0.21 &                     No \\
                filler\_words\_p1 vs filler\_words\_p2 &            3.10 &                       Yes &         0.54 &                     No \\
filler\_words\_percent\_p1 vs filler\_words\_percent\_p2 &            3.03 &                       Yes &         0.10 &                     No \\
                    weak\_words\_p1 vs weak\_words\_p2 &            3.07 &                       Yes &         0.86 &                     No \\
    weak\_words\_percent\_p1 vs weak\_words\_percent\_p2 &            2.78 &                       Yes &         0.59 &                     No \\
  eye\_contact\_percent\_p1 vs eye\_contact\_percent\_p2 &           -1.05 &                        No &        -1.38 &                     No \\
            pacing\_w\_per\_m\_p1 vs pacing\_w\_per\_m\_p2 &            0.68 &                        No &        -0.51 &                     No \\
                    pacing\_low\_p1 vs pacing\_low\_p2 &            1.43 &                        No &        -1.23 &                     No \\
                  pacing\_high\_p1 vs pacing\_high\_p2 &            2.07 &                        No &         0.41 &                     No \\
        pacing\_variation\_p1 vs pacing\_variation\_p2 &            0.78 &                        No &         1.01 &                     No \\
                      pause\_min\_p1 vs pause\_min\_p2 &           -0.34 &                        No &         1.12 &                     No \\
                      pause\_max\_p1 vs pause\_max\_p2 &           -0.77 &                        No &         0.29 &                     No \\
                    pause\_mean\_p1 vs pause\_mean\_p2 &           -0.65 &                        No &         0.62 &                     No \\
    Confidence\_Overall\_p1 vs Confidence\_Overall\_p2 &           -5.91 &                       Yes &        -5.49 &                    Yes \\
Confidence\_Repetition\_p1 vs Confidence\_Repetiti... &           -0.64 &                        No &        -1.23 &                     No \\
Confidence\_Fluency\_p1 vs Confidence\_F... &           -1.15 &                        No &        -1.15 &                     No \\
Confidence\_Pace\_p1 vs Confidence\_Pa... &           -2.17 &                       Yes &        -2.62 &                    Yes \\
  Confidence\_Loudness\_p1 vs Confidence\_Loudness\_p2 &            0.29 &                        No &        -2.88 &                    Yes \\
Confidence\_EyeConfidence\_ontaConfidence\_t\_p1 vs... &           -2.38 &                       Yes &        -4.76 &                    Yes \\
      Confidence\_Pauses\_p1 vs Confidence\_Pauses\_p2 &           -2.96 &                       Yes &        -1.73 &                     No \\
        Confidence\_Smile\_p1 vs Confidence\_Smile\_p2 &           -2.44 &                       Yes &        -1.47 &                     No \\
Confidence\_Pronunciation\_p1 vs Confid... &           -1.72 &                        No &        -3.16 &                    Yes \\
Satisfaction\_Overall\_p1 vs Satisfaction\_Overall\_p2 &           -7.64 &                       Yes &        -3.37 &                    Yes \\
Satisfaction\_Repetition\_p1 vs Satisfaction\_Repe... &           -2.75 &                       Yes &        -2.06 &                     No \\
Satisfaction\_Fluency\_p1 vs Satisfaction\_Fluency\_p2 &           -2.61 &                       Yes &        -2.35 &                    Yes \\
  Satisfaction\_Pacing\_p1 vs Satisfaction\_Pacing\_p2 &           -3.76 &                       Yes &        -2.38 &                    Yes \\
Satisfaction\_Loudness\_p... &           -0.29 &                        No &        -0.71 &                     No \\
Satisfaction\_EyeContact\_p1 vs Satisfaction\_EyeC... &           -2.44 &                       Yes &        -3.04 &                    Yes \\
Satisfaction\_Pause\_p1 ... &           -4.43 &                       Yes &        -2.90 &                    Yes \\
Satisfaction\_Smile\_p1 vs Satisfacti... &           -1.42 &                        No &        -1.47 &                     No \\
Satisfaction\_Pronunciation\_p1 vs Satisfaction\_P... &           -1.00 &                        No &        -4.32 &                    Yes \\
                          Express\_p1 vs Express\_p2 &           -3.39 &                       Yes &        -4.37 &                    Yes \\
              Communication\_p1 vs Communication\_p2 &           -2.48 &                       Yes &        -0.76 &                     No \\
\bottomrule
\end{tabular}
\caption{Paired T-Test For AI-Clone and Self-Recording Groups 
\label{Paired-t-test-CE}}
\end{table}

\begin{table}[H]
\centering
\begin{tabular}{lcccc}
\toprule
{} &  T(Promotion) & Significance  &  T(Prevention) & Significance \\
\midrule
word\_count\_p1 vs word\_count\_p2                     &             1.495603 &                       No &              2.069154 &                        No \\
repetition\_p1 vs repetition\_p2                     &             1.963961 &                       No &              2.566756 &                       Yes \\
repetition\_percent\_p1 vs repetition\_percent\_p2     &             0.239229 &                       No &              1.857143 &                        No \\
filler\_words\_p1 vs filler\_words\_p2                 &             5.280648 &                      Yes &              3.572379 &                       Yes \\
filler\_words\_percent\_p1 vs filler\_words\_percent\_p2 &             3.239338 &                      Yes &              3.126775 &                       Yes \\
weak\_words\_p1 vs weak\_words\_p2                     &             1.980208 &                       No &              4.060835 &                       Yes \\
weak\_words\_percent\_p1 vs weak\_words\_percent\_p2     &             1.746130 &                       No &              3.536934 &                       Yes \\
eye\_contact\_percent\_p1 vs eye\_contact\_percent\_p2   &            -1.666187 &                       No &              1.160104 &                        No \\
pacing\_w\_per\_m\_p1 vs pacing\_w\_per\_m\_p2             &             0.017445 &                       No &              0.653331 &                        No \\
pacing\_low\_p1 vs pacing\_low\_p2                     &            -0.026353 &                       No &              0.148853 &                        No \\
pacing\_high\_p1 vs pacing\_high\_p2                   &             0.268635 &                       No &              1.270466 &                        No \\
pacing\_variation\_p1 vs pacing\_variation\_p2         &             0.325423 &                       No &              0.740699 &                        No \\
pause\_min\_p1 vs pause\_min\_p2                       &             0.413899 &                       No &              0.437498 &                        No \\
pause\_max\_p1 vs pause\_max\_p2                       &             0.563535 &                       No &             -0.490012 &                        No \\
pause\_mean\_p1 vs pause\_mean\_p2                     &             0.715467 &                       No &             -0.286169 &                        No \\
Confidence\_Overall\_p1 vs Confidence\_Overall\_p2     &            -5.870395 &                      Yes &             -1.718466 &                        No \\
Confidence\_Repetition\_p1 vs Confidence\_Repetiti... &            -2.846974 &                      Yes &             -0.263589 &                        No \\
Confidence\_Flueny\_p1 vs Confidence\_Flueny\_p2       &            -1.585356 &                       No &             -2.958040 &                       Yes \\
Confidence\_Pace\_p1 vs Confidence\_Pace\_p2           &            -1.581139 &                       No &             -2.673651 &                       Yes \\
Confidence\_Loudness\_p1 vs Confidence\_Loudness\_p2   &            -2.085549 &                       No &             -1.381699 &                        No \\
Confidence\_EyeContact\_p1 vs Confidence\_EyeConta... &            -2.442292 &                      Yes &             -1.457256 &                        No \\
Confidence\_Pauses\_p1 vs Confidence\_Pauses\_p2       &            -3.055050 &                      Yes &             -1.975658 &                        No \\
Confidence\_Smile\_p1 vs Confidence\_Smile\_p2         &            -1.657850 &                       No &             -1.748646 &                        No \\
Confidence\_Pronunciation\_p1 vs Confidence\_Pronu... &            -3.666494 &                      Yes &             -0.211154 &                        No \\
Satisfaction\_Overall\_p1 vs Satisfaction\_Overall\_p2 &            -6.204837 &                      Yes &             -3.500000 &                       Yes \\
Satisfaction\_Repetition\_p1 vs Satisfaction\_Repe... &            -3.089572 &                      Yes &             -2.442292 &                       Yes \\
Satisfaction\_Fluency\_p1 vs Satisfaction\_Fluency\_p2 &            -4.063180 &                      Yes &             -1.870829 &                        No \\
Satisfaction\_Pace\_p1 vs Satisfaction\_Pace\_p2       &            -3.095624 &                      Yes &             -3.378463 &                       Yes \\
Satisfaction\_Loudness\_p1 vs Satisfaction\_Loudne... &            -2.085549 &                       No &             -1.701926 &                        No \\
Satisfaction\_EyeContact\_p1 vs Satisfaction\_EyeC... &            -3.162278 &                      Yes &             -3.055050 &                       Yes \\
Satisfaction\_Pause\_p1 vs Satisfaction\_Pause\_p2     &            -3.240370 &                      Yes &             -2.449490 &                       Yes \\
Satisfaction\_Smile\_p1 vs Satisfaction\_Smile\_p2     &            -0.844616 &                       No &             -1.895158 &                        No \\
Satisfaction\_Pronunciation\_p1 vs Satisfaction\_P... &            -2.954989 &                      Yes &             -0.564076 &                        No \\
Express\_p1 vs Express\_p2                           &            -2.322693 &                      Yes &             -3.674235 &                       Yes \\
Communication\_p1 vs Communication\_p2               &            -0.487122 &                       No &             -2.256304 &                       Yes \\
\bottomrule
\end{tabular}
\caption{Paired T-Test For Promotion and Prevention Groups}
\label{Paired-t-test-PP}
\end{table}

\begin{table}[H]
\centering
\begin{tabular}{lrrrrr}
\hline
 index                                            &   F\_value &   p\_value &   df\_group &   df\_error &   eta\_squared \\
\hline
 word\_count\_p2                                    &    0.0152 &    0.9027 &          1 &         26 &        0.0003 \\
 repetition\_p2                                    &    0.0667 &    0.7982 &          1 &         26 &        0.0025 \\
 repetition\_percent\_p2                            &    0.1766 &    0.6778 &          1 &         26 &        0.0048 \\
 filler\_words\_p2                                  &    2.2560 &    0.1451 &          1 &         26 &        0.0563 \\
 filler\_words\_percent\_p2                          &    3.3421 &    0.0790 &          1 &         26 &        0.0763 \\
 weak\_words\_p2                                    &    1.1586 &    0.2916 &          1 &         26 &        0.0283 \\
 weak\_words\_percent\_p2                            &    1.0889 &    0.3063 &          1 &         26 &        0.0310 \\
 eye\_contact\_percent\_p2                           &    0.6808 &    0.4168 &          1 &         26 &        0.0073 \\
 pacing\_w\_per\_m\_p2                                &    1.2174 &    0.2800 &          1 &         26 &        0.0067 \\
 pacing\_low\_p2                                    &    4.4982 &    0.0436 &          1 &         26 &        0.0390 \\
 pacing\_high\_p2                                   &    1.5016 &    0.2314 &          1 &         26 &        0.0173 \\
 pacing\_variation\_p2                              &    0.3609 &    0.5532 &          1 &         26 &        0.0108 \\
 pause\_min\_p2                                     &    1.0610 &    0.3125 &          1 &         26 &        0.0061 \\
 pause\_max\_p2                                     &    0.5418 &    0.4683 &          1 &         26 &        0.0072 \\
 pause\_mean\_p2                                    &    0.7247 &    0.4024 &          1 &         26 &        0.0047 \\
 Confidence\_Overall\_p2                            &    0.1133 &    0.7391 &          1 &         26 &        0.0025 \\
 Confidence\_Repetition\_p2                         &    1.8943 &    0.1805 &          1 &         26 &        0.0670 \\
 Confidence\_Fluency\_p2                  &    0.3840 &    0.5409 &          1 &         26 &        0.0130 \\
 Confidence\_Pace\_p2                   &    0.1076 &    0.7456 &          1 &         26 &        0.0041 \\
 Confidence\_Loudness\_p2                           &    2.6903 &    0.1130 &          1 &         26 &        0.0689 \\
 Confidence\_EyeContact\_p2     &    0.3810 &    0.5424 &          1 &         26 &        0.0087 \\
 Confidence\_Pauses\_p2                             &    0.0325 &    0.8584 &          1 &         26 &        0.0010 \\
 Confidence\_Smile\_p2                              &    0.2277 &    0.6372 &          1 &         26 &        0.0072 \\
 Confidence\_Pronunciation\_p2            &    2.6962 &    0.1126 &          1 &         26 &        0.0737 \\
 Satisfaction\_Overall\_p2                          &    0.0121 &    0.9134 &          1 &         26 &        0.0003 \\
 Satisfaction\_Repetition\_p2                       &    1.8033 &    0.1909 &          1 &         26 &        0.0551 \\
 Satisfaction\_Fluency\_p2                          &    0.9456 &    0.3398 &          1 &         26 &        0.0322 \\
 Satisfaction\_Pacing\_p2                           &    0.0877 &    0.7694 &          1 &         26 &        0.0027 \\
 Satisfaction\_Loudness\_p2 &    0.0186 &    0.8926 &          1 &         26 &        0.0003 \\
 Satisfaction\_EyeContact\_p2                       &    1.4508 &    0.2393 &          1 &         26 &        0.0405 \\
 Satisfaction\_Pause\_\_p2   &    0.0087 &    0.9265 &          1 &         26 &        0.0003 \\
 Satisfaction\_Smile\_p2                &    3.5334 &    0.0714 &          1 &         26 &        0.1195 \\
 Satisfaction\_Pronunciation\_p2                    &   10.4976 &    0.0033 &          1 &         26 &        0.2566 \\
 Express\_p2                                       &    1.2524 &    0.2733 &          1 &         26 &        0.0295 \\
 Communication\_p2                                 &    0.1626 &    0.6901 &          1 &         26 &        0.0046 \\
\hline
\end{tabular}
\caption{ANCOVA Result For Self-Recording and AI-Clone Groups 
\label{Ancova-CE}}
\end{table}

\begin{table}[H]
\centering
\begin{tabular}{lrrrrr}
\toprule
           Dependent Variable &  F Value &  p Value &  df Group &  df Error &  Eta Squared \\
\midrule
                word\_count\_p2 &    0.132 &    0.719 &     1.000 &    27.000 &        0.001 \\
                repetition\_p2 &    0.132 &    0.720 &     1.000 &    27.000 &        0.004 \\
        repetition\_percent\_p2 &    0.239 &    0.629 &     1.000 &    27.000 &        0.006 \\
              filler\_words\_p2 &    1.353 &    0.255 &     1.000 &    27.000 &        0.010 \\
      filler\_words\_percent\_p2 &    1.567 &    0.221 &     1.000 &    27.000 &        0.018 \\
                weak\_words\_p2 &    0.218 &    0.644 &     1.000 &    27.000 &        0.003 \\
        weak\_words\_percent\_p2 &    0.088 &    0.768 &     1.000 &    27.000 &        0.002 \\
       eye\_contact\_percent\_p2 &    1.942 &    0.175 &     1.000 &    27.000 &        0.031 \\
            pacing\_w\_per\_m\_p2 &    0.192 &    0.665 &     1.000 &    27.000 &        0.001 \\
                pacing\_low\_p2 &    0.005 &    0.942 &     1.000 &    27.000 &        0.000 \\
               pacing\_high\_p2 &    0.440 &    0.513 &     1.000 &    27.000 &        0.003 \\
          pacing\_variation\_p2 &    0.706 &    0.408 &     1.000 &    27.000 &        0.020 \\
                 pause\_min\_p2 &    0.003 &    0.956 &     1.000 &    27.000 &        0.000 \\
                 pause\_max\_p2 &    0.570 &    0.457 &     1.000 &    27.000 &        0.011 \\
                pause\_mean\_p2 &    0.492 &    0.489 &     1.000 &    27.000 &        0.004 \\
        Confidence\_Overall\_p2 &    0.918 &    0.346 &     1.000 &    27.000 &        0.022 \\
     Confidence\_Repetition\_p2 &    1.387 &    0.249 &     1.000 &    27.000 &        0.048 \\
         Confidence\_Flueny\_p2 &    0.159 &    0.693 &     1.000 &    27.000 &        0.005 \\
           Confidence\_Pace\_p2 &    0.278 &    0.602 &     1.000 &    27.000 &        0.010 \\
       Confidence\_Loudness\_p2 &    0.040 &    0.843 &     1.000 &    27.000 &        0.001 \\
     Confidence\_EyeContact\_p2 &    0.258 &    0.615 &     1.000 &    27.000 &        0.007 \\
         Confidence\_Pauses\_p2 &    0.053 &    0.820 &     1.000 &    27.000 &        0.001 \\
          Confidence\_Smile\_p2 &    0.109 &    0.744 &     1.000 &    27.000 &        0.003 \\
  Confidence\_Pronunciation\_p2 &    5.434 &    0.027 &     1.000 &    27.000 &        0.101 \\
      Satisfaction\_Overall\_p2 &    1.770 &    0.195 &     1.000 &    27.000 &        0.036 \\
   Satisfaction\_Repetition\_p2 &    0.061 &    0.807 &     1.000 &    27.000 &        0.001 \\
      Satisfaction\_Fluency\_p2 &    2.348 &    0.137 &     1.000 &    27.000 &        0.063 \\
         Satisfaction\_Pace\_p2 &    0.349 &    0.559 &     1.000 &    27.000 &        0.009 \\
     Satisfaction\_Loudness\_p2 &    0.358 &    0.555 &     1.000 &    27.000 &        0.006 \\
   Satisfaction\_EyeContact\_p2 &    0.060 &    0.808 &     1.000 &    27.000 &        0.001 \\
        Satisfaction\_Pause\_p2 &    0.052 &    0.821 &     1.000 &    27.000 &        0.002 \\
        Satisfaction\_Smile\_p2 &    0.140 &    0.711 &     1.000 &    27.000 &        0.004 \\
Satisfaction\_Pronunciation\_p2 &    3.044 &    0.092 &     1.000 &    27.000 &        0.057 \\
                   Express\_p2 &    0.060 &    0.809 &     1.000 &    27.000 &        0.001 \\
             Communication\_p2 &    0.964 &    0.335 &     1.000 &    27.000 &        0.020 \\
\bottomrule
\end{tabular}
\caption{ANCOVA Result For Promotion and Prevention Groups 
\label{Ancova-PP}}
\end{table}

\section{Technology Drawbacks and Ethical Concerns}\label{ethics}
Below we also report the concerns from participants' interviews.

\paragraph{\textbf{\textit{Personal identity and self-hood are of utmost importance to make sure their safety use. }}} The central concern related to "self" underscores the apprehension regarding AI-clones potentially "dehumanizing" the individuals they replicate. This fear is not unfounded, as participants explicitly voiced worries over their personal identity being misappropriated or used without their consent. For instance, one participant emphasized the need for autonomy and security by stating,\textit{ "I want to make sure that my AI-clone will not be used elsewhere without me knowing that (P29)."}

Moreover, the desire for control over one's digital representation was evident, as another participant expressed a need for a mechanism to distinguish their AI-clone from their actual self, highlighting a nuanced approach to identity management in the digital age. They remarked,  "I want a a package myself. Although AI is mirroring me, but I want a way to deliberately manipulate those differences so no one can tell if this is my AI-clone besides myself (P4).” These responses reveal a deep-seated concern for personal agency and the integrity of one's identity in the face of advancing AI technology. They reflect a collective desire among participants to navigate the delicate balance between leveraging AI for personal or professional gain and safeguarding their individuality against potential misuse or exploitation.

\paragraph{\textbf{AI-clone's advanced language enhances learning, but excessive complexity can reduce engagement}} When comparing the AI-clones with the advanced vocabulary, participants noted the AI's advanced version facilitated learning through its use of an expanded vocabulary. For instance, one participant observed, \textit{"It used more expanded vocabulary. You know like this avatar is speaking, unscripted. For example, "soothing". “I am aware of that word, but I didn't really think of using that word during my presentation. I felt like that skill is really particular to a native speaker. You know just in importing the other words. I'm not sure if that's the right expression. The AI was less academic, more soothing, and more caring, which felt contextually appropriate. I think I used this word in my second presentation (P7)."}

Another participant shared a different perspective: \textit{"Hearing AI's speech, I realize that there should be a story! I am like an audience listening to myself speaking. The AI is based on me, but it is so good that I feel it is another person. I feel like I can see my speech from a different person. This is very similar to me, and I think this is very important. If a person is different from me I will think this is not my style. Each person has different characteristics. We have to consider my true level of personality, and tones (P11)."} These insights highlight a delicate balance between utilizing advanced language for educational purposes and ensuring it doesn't overwhelm or alienate learners by straying too far from their personal expression and comfort level.

\paragraph{\textbf{\textit{Will it Represent Me? Controlling self-clone identities to foster alignments in human-AI interactions}}}

The AI-clone is a standard version and should allow for scenario customization.  \textit{"I want to control the ambient of the environment. I want to adjust the tone, to a more enlightening, more serious tone. Sometimes there’s a serious presentation, it depends on the situation. Sometimes you need less smile, even with your mouth closed. Don’t show your teeth at all. Have some templates so we can choose, e.g., this is for a school interview, this is for a job interview, so you can professional advice (P3). "} This sentiment underscores a deeper concern among participants about the potential of AI-clones to accurately and consistently represent their human counterparts in various settings. The desire for scenario-specific customization reflects a need for authenticity and appropriate presentation, balancing professionalism with personal expression.

Moreover, the expectation that AI-clones should evolve over time to mirror changes in the individual's appearance and preferences speaks to a broader anticipation of AI as dynamic entities capable of growth.  For example,  \textit{"I think I want to make it prettier, wearing make-up because sometimes I don't really want to do that in an online presentation. But you know, it can't feel too good to be true. Also, if my hair style changed, it should change as well (P41)." } These insights reveal a complex relationship between individuals and their AI counterparts, highlighting a keen interest in ensuring these digital clones do not merely mimic but genuinely represent the person's evolving identity. This approach not only addresses concerns of authenticity and trust but also reflects a nuanced understanding of the role AI can play in extending personal identity into digital spaces.

\paragraph{\textbf{\textit{
Navigating the complexities of AI-clones in AI-mediated interpersonal relationships}}}
Participants raised concerns about the potential for AI-clones to disrupt interpersonal relationships, especially when nuances of personal interaction are overlooked or when individuals are unaware they're engaging with an AI-clone. This issue is particularly pronounced in contexts where AI-clones are utilized for specific training purposes but may not fully encapsulate the broad spectrum of human communication. For example, the lack of capturing the subtleties in interpersonal dynamics by AI-clones can lead to misunderstandings. As some participants noted, \textit{"An overly positive version of me (P3)"}, and \textit{"Sometimes they don't blink and they are always smiling in this official expression. I want myself to be viewed as a "well-educated" person, but I need a way to tell AI what do I perceive "well-educated" is (P6)."}

Moreover, participants worry about presenting a consistent image across different platforms and situations. \textit{"I am worried if I only use it for some presentation, but not all aspects of my speech and language, will my presentation leave an impression to the audience that is not consistent with who I really am? Similarly, if my AI alone is shared with others, it will also have this issue (P36). "} These insights suggest that while AI-clones offer innovative opportunities for personal development and training, there's a critical need for these technologies to more accurately reflect the individual's personality and communication style. Ensuring that users and their audiences are aware of when they are interacting with AI-clones is vital for maintaining trust and authenticity in interpersonal relationships.

\end{document}